\documentclass[reprint,
 amsmath,amssymb,
 aps,
floatfix,
showkeys
]{revtex4-1}

\usepackage[USenglish]{babel}
\usepackage{slashed}
\usepackage{physics}
\usepackage{graphicx}
\usepackage{dcolumn}
\usepackage{bm}
\usepackage[usenames]{color}
\usepackage{xcolor}
\usepackage{float}

\newcommand{\be}{\begin{equation}}
\newcommand{\ee}{\end{equation}}
\newcommand{\bea}{\begin{eqnarray}}
\newcommand{\eea}{\end{eqnarray}}
\newcommand{\beas}{\begin{eqnarray*}}
\newcommand{\eeas}{\end{eqnarray*}}

\begin{document}

\title{Magnetic corrections to the boson self-coupling and boson-fermion coupling in the linear sigma model with quarks}


\author{Alejandro Ayala$^{1,2}$, Jos\'e Luis Hern\'andez$^{1}$, L. A. Hern\'andez$^{1,2,3}$, Ricardo L. S. Farias$^4$, R. Zamora$^{5,6}$ }
\affiliation{%
$^1$Instituto de Ciencias Nucleares, Universidad Nacional Aut\'onoma de M\'exico, Apartado Postal 70-543, CdMx 04510, Mexico.\\
$^2$Centre for Theoretical and Mathematical Physics, and Department of Physics, University of Cape Town, Rondebosch 7700, South Africa.\\
$^3$Facultad de Ciencias de la Educaci\'on, Universidad Aut\'onoma de Tlaxcala, Tlaxcala, 90000, Mexico.\\
$^4$Departamento de F\'isica, Universidade Federal de Santa Maria, Santa Maria, RS 97105-900, Brazil.\\
$^5$Instituto de Ciencias B\'asicas, Universidad Diego Portales, Casilla 298-V, Santiago, Chile.\\
$^6$Centro de Investigaci\'on y Desarrollo en Ciencias Aeroespaciales (CIDCA), Fuerza A\'erea de Chile,  Santiago, Chile.}%


\begin{abstract}

We compute the magnetic field-induced modifications to the boson self-coupling and the boson-fermion coupling, in the static limit, using an effective model of QCD, the linear sigma model with quarks. The former is computed for arbitrary field strengths as well as using the strong field approximation. The latter is obtained in the strong field limit. The arbitrary field result for the boson self-coupling depends on the ultraviolet renormalization scale and this dependence cannot be removed by a simple vacuum subtraction. Using the strong field result as a guide, we find the appropriate choice for this scale and discuss the physical implications. The boson-fermion coupling depends on the Schwinger's phase and we show how this phase can be treated consistently in such a way that the magnetic field induced vertex modification is both gauge invariant and can be written with an explicit factor corresponding to energy-momentum conservation for the external particles. Both couplings show a modest decrease with the field strength.  

\end{abstract}

\keywords{Quantum Chromodynamics, Linear Sigma Model with Quarks, Magnetic Fields}

\maketitle

\section{Introduction}\label{sec1}

The effects of magnetic fields on the properties of strongly interacting matter have gathered a great deal of interest over the last years. The main driving motivation is the lattice QCD (LQCD) discovery of the inverse magnetic catalysis (IMC) phenomenon~\cite{LQCD}, whereby for temperatures above the chiral restoration one, the quark-antiquark condensate decreases and the chiral restoration temperature itself also decreases, as a function of the field intensity. The origin of the IMC has been intensively studied, see for example Refs.~\cite{Bruckmann,Farias,Ferreira,Ayala0, Ayala1,Ayala2,Ayala3,Avancini,Ayala4,vertex1,vertex2,qcdcoupling,Mueller,imcreview}. 

In addition, much effort has also been devoted to study the basic properties of magnetized hadronic degrees of freedom. The subject is important {\it e.g.} for systems such as cold neutron stars and heavy-ion collisions. As is well known, the nuclear equation of state is affected by baryon and mesons masses and couplings which motivates studies aimed to understand how these parameters change in the presence of electromagnetic fields~\cite{bali01, iranianos, simonov03,aguirre02,tetsuya,dudal04,kevin,gubler, noronha01,morita,morita02,sarkar03,band}. Different  effective QCD models~\cite{Ayalachi,nosso1,nosso03,zhuang,iran,scoccola01,huang01,scoccola02,scoccola03,luch,farias01,mao01, sarkar,sarkar02, zhang01,huan02,simonov01,fraga01,aguirre,taya,shinya,andersen01,kojo,simonov02,ghoshrho,AMMmesons,TBspectralprop,ghoshEPJA}, LQCD simulations~\cite{luschev,luschv02,luschv03,bali02,bali03,hidaka,Ding2} as well as holografic QCD models~\cite{Avila,Avila:2020ved,dudal,dudal02} have been used to describe the behavior of light meson masses. More recently, efforts have also been carried out to describe the behavior of light baryons in the presence of  magnetic fields~\cite{andrei02,he,nucleon,barionslattice}. 
In particular, the recent LQCD results for the magnetic field-driven modifications of neutral and charged mesons show that the neutral pion mass monotonically decreases, whereas the mass of the charged pions monotonically increases, both as functions of the field intensity~\cite{bali02,Ding}. The former cannot be fully reproduced by calculations within effective models that do not consider accounting for magnetic field modifications of the couplings~\cite{Das,Li}.  

When the linear sigma model with quarks (LSMq) is used as an effective QCD model, it has been shown that the IMC can be reasonably well described when temperature, as well as magnetic field corrections, are incorporated into self-energies and couplings~\cite{Ayala1,Ayala3}. The decreasing of the neutral pion mass with the magnetic field strength can also be understood when in the weak field limit the meson self-coupling is dressed to include magnetic field effects~\cite{pionmassmag}. In order to find out whether or not in the LSMq, the behaviour of the pion masses can be described over a wider range of magnetic field intensities, it is important to compute the magnetic field-induced corrections to the interaction vertices.

In this work, we address this question and compute the one-loop magnetic field corrections to the boson self-coupling and the boson-fermion coupling in the LSMq. In doing so, we address some important details involving the effects introduced by the renormalization scale as well as those introduced by the Schwinger's phase in calculations involving three particles propagating within loops in the presence of magnetic fields. The work is organized as follows: In Sec.~\ref{secII} we introduce the LSMq writing the Lagrangian in terms of the charged pion degrees of freedom and including an explicit symmetry breaking term. In Sec.~\ref{secIII} we recall the way the magnetic field effects are introduced for the propagators of charged bosons and fermions. In Sec.~\ref{secIV} we compute the modification to the boson self-coupling in the presence of a magnetic field. We show that the modification depends on the renormalization scale and that for this to match the result obtained in the strong field limit, one needs to resort to a suitable choice for this scale. In Sec.~\ref{secV} we compute the magnetic field induced modification to the boson-fermion coupling and discuss in detail the effect of the Schwinger's phase. We show that this leads to a plausible result respecting energy-momentum conservation for the external particles when these are described as plane waves and thus when we neglect propagation over large space-time intervals.  Finally, we summarize and provide an outlook of our results in Sec.~\ref{concl} leaving for the appendices the details of the calculation of the boson self-coupling and the boson-fermion coupling.

\section{The LSMq}\label{secII}

The LSMq is an effective theory that captures the approximate chiral symmetry of QCD. It describes the interactions among small-mass mesons and quarks. We work with a Lagrangian invariant under $SU(2)_{L}\times SU(2)_{R}$ chiral transformations
\begin{eqnarray}
\mathcal{L}&=&\frac{1}{2}(\partial_{\mu}\sigma)^{2}+\frac{1}{2}(\partial_{\mu}\vec{\pi})^{2}+\frac{a^{2}}{2}(\sigma^{2}+\vec{\pi}^{2})-\frac{\lambda}{4}(\sigma^{2}
+\vec{\pi}^{2})^{2}\nonumber\\
&+&i\bar{\psi}\gamma^{\mu}\partial_{\mu}\psi-ig\gamma^{5}\bar{\psi} \vec{\tau} \cdot \vec{\pi}\psi-g\bar{\psi}\psi\sigma,
\label{sigmalinearmodellagrangian}
\end{eqnarray}
where  
$\vec{\tau}=(\tau_{1},\tau_{2},\tau_{3})$
are the Pauli matrices, \begin{eqnarray}
\psi_{L,R}= \begin{pmatrix} u \\ d \end{pmatrix}_{L,R},
\label{doublet}
\end{eqnarray}is a  $SU(2)_{L,R}$ doublet, $\sigma$ is a real scalar field and $\vec{\pi}=(\pi_{1},\pi_{2},\pi_{3})$ is a triplet of real scalar fields. $\pi_3$ corresponds to the neutral pion whereas the charged ones are represented by the combinations
\begin{equation}
    \pi_{-}=\frac{1}{\sqrt{2}}(\pi_{1}+i\pi_{2}), \quad \pi_{+}=\frac{1}{\sqrt{2}}(\pi_{1}-i\pi_{2}).
\end{equation}
$\lambda$ is the boson's self-coupling and $g$ is the fermion-boson coupling. $a^2>0$ is the mass parameter. Eq.~(\ref{sigmalinearmodellagrangian}) can be written in terms of the charged and neutral-pion degrees of freedom as
\begin{eqnarray}
\mathcal{L}&=&\frac{1}{2}[(\partial_{\mu}\sigma)^{2}+(\partial_{\mu}\pi_{0})^{2}] +\partial_{\mu}\pi_{-}\partial^{\mu}\pi_{+}+\frac{a^{2}}{2}(\sigma^{2}+\pi_{0}^{2})\nonumber\\
&+&a^{2}\pi_{-}\pi_{+} -\frac{\lambda}{4}(\sigma^{4}+4\sigma^{2}\pi_{-}\pi_{+}+2\sigma^{2}\pi_{0}+4\pi_{-}^{2}\pi_{+}^{2}\nonumber\\
&+&4\pi_{-}\pi_{+}\pi_{0}^{2}+\pi_{0}^{4})+i\bar{\psi}\slashed{\partial}\psi-g\bar{\psi}\psi\sigma-ig\gamma^{5}\bar{\psi}(\tau_{+}\pi_{+}\nonumber\\
&+&\tau_{-}\pi_{-}+\tau_{3}\pi_{0})\psi,
\label{sigmalinearmodellagrangianmodified}    
\end{eqnarray}
where we introduced the combination of Pauli matrices
\begin{eqnarray}
    \tau_{+}=\frac{1}{\sqrt{2}}(\tau_{1}+i\tau_{2}), \quad \tau_{-}=\frac{1}{\sqrt{2}}(\tau_{1}-i\tau_{2}).
\end{eqnarray}
After chiral symmetry is spontaneously broken, the field $\sigma$ acquires a non-vanishing vacuum expectation value $\sigma\to\sigma + v$, which breaks the $SU(2)_{L}\times SU(2)_{R}$ symmetry down to $SU(2)_{L+R}$,  resulting in the Lagrangian
\begin{eqnarray}
    \mathcal{L}&=&\frac{1}{2}\partial_{\mu}\sigma \partial^{\mu}\sigma+\frac{1}{2}\partial_{\mu}\pi_{0}\partial^{\mu}\pi_{0}+\partial_{\mu}\pi_{-}\partial^{\mu}\pi_{+}\nonumber\\
    &-&\frac{1}{2}m_{\sigma}^{2}\sigma^{2}-\frac{1}{2}m_{\pi}^{2}\pi_{0}^{2}-m_{\pi}^{2}\pi_{-}\pi_{+}+i\bar{\psi}\slashed{\partial}\psi\nonumber\\
    &-&m_{f}\bar{\psi}\psi+\mathcal{L}_{int}-V_{tree},
    \label{linearsigmamodelSSB}
\end{eqnarray}
where the interaction Lagrangian is defined as
\begin{equation}
\begin{split}
    \mathcal{L}_{int}&=-\frac{\lambda}{4}\sigma^{4}-\lambda v\sigma^{3}-\lambda v^{3}\sigma-\lambda\sigma^{2}\pi_{-}\pi_{+} -2\lambda v \sigma\pi_{-}\pi_{+}\\
    &-\frac{\lambda}{2}\sigma^{2}\pi_{0}^{2}-\lambda v\sigma \pi_{0}^{2}-\lambda \pi_{-}^{2}\pi_{+}^{2}-\lambda\pi_{-}\pi_{+}\pi_{0}^{2}-\frac{\lambda}{4}\pi_{0}^{4}\\ 
    &+a^{2}v\sigma -g\bar{\psi}\psi\sigma-ig\gamma^{5}\bar{\psi}\left(\tau_{+}\pi_{+}+\tau_{-}\pi_{-}+\tau_{3}\pi_{0}\right)\psi,
    \label{interactinglagrangian}
\end{split}    
\end{equation}
and the tree level potential can be expressed as
\begin{eqnarray}
    V_{tree}=-\frac{a^{2}}{2}v^{2}+\frac{\lambda}{4}v^4.
\label{treeeffectivepotential}
\end{eqnarray}
As can be seen from Eqs.~\eqref{linearsigmamodelSSB},~\eqref{interactinglagrangian} and~\eqref{treeeffectivepotential} there are new terms which depend on  $v$. In particular, the fields develop dynamic masses given by
\begin{equation}
     m_{\sigma}^{2}=3\lambda v^2-a^2, \quad m_{\pi}^{2}=\lambda v^2-a^2, \quad m_{f}=gv.
\label{masses}
\end{equation}
The tree level potential develops a minimum, called the vacuum expectation value of the $\sigma$ field, namely 
\begin{equation}
    v_{0}=\sqrt{\frac{a^{2}}{\lambda}}.
\label{vev}
\end{equation}
Notice that when $v=v_0$, the linear term in $\sigma$ vanishes and the pions become massless. However, the $\sigma$ and quark fields remain massive.

In order to include a finite vacuum pion mass, one adds an explicit symmetry breaking term in the Lagrangian of Eq.~\eqref{linearsigmamodelSSB} such that
\begin{equation}
    \mathcal{L}\rightarrow \mathcal{L'}=\mathcal{L}+\frac{m_{\pi}^{2}}{2}v(\sigma+v).
\end{equation}
This term modifies the tree-level potential. In particular, the minimum is shifted such that
\begin{equation}
    v_{0}\rightarrow v_{0}'=\sqrt{\frac{a^{2}+m_{\pi}^{2}}{\lambda}}.
\end{equation}
Correspondingly, the expressions for the masses, evaluated at the minimum obtained after the explicit breaking of the symmetry, are given by 
\begin{eqnarray}
    &&m_{f}(v_{0}')=g\sqrt{\frac{a^{2}+m_{\pi}^{2}}{\lambda}},\nonumber\\
    &&m_{\sigma}^2(v_{0}')=2a^{2}+3m_{\pi}^{2}, \nonumber\\
    &&m_{\pi}^2(v_{0}')=m_{\pi}^{2}.
\end{eqnarray}
Furthermore, from Eq.~\eqref{masses}, we can get an expression for the parameter $a$, which is given by
\begin{equation}
    a=\sqrt{\frac{m_{\sigma}^{2}-3m_{\pi}^{2}}{2}}.
\end{equation}
Setting $m_{\pi}= 140$ MeV and $m_{\sigma}= 400-600$ MeV, we get $a= 225-390$ MeV.
\begin{figure}[t]
    \includegraphics[scale=0.24]{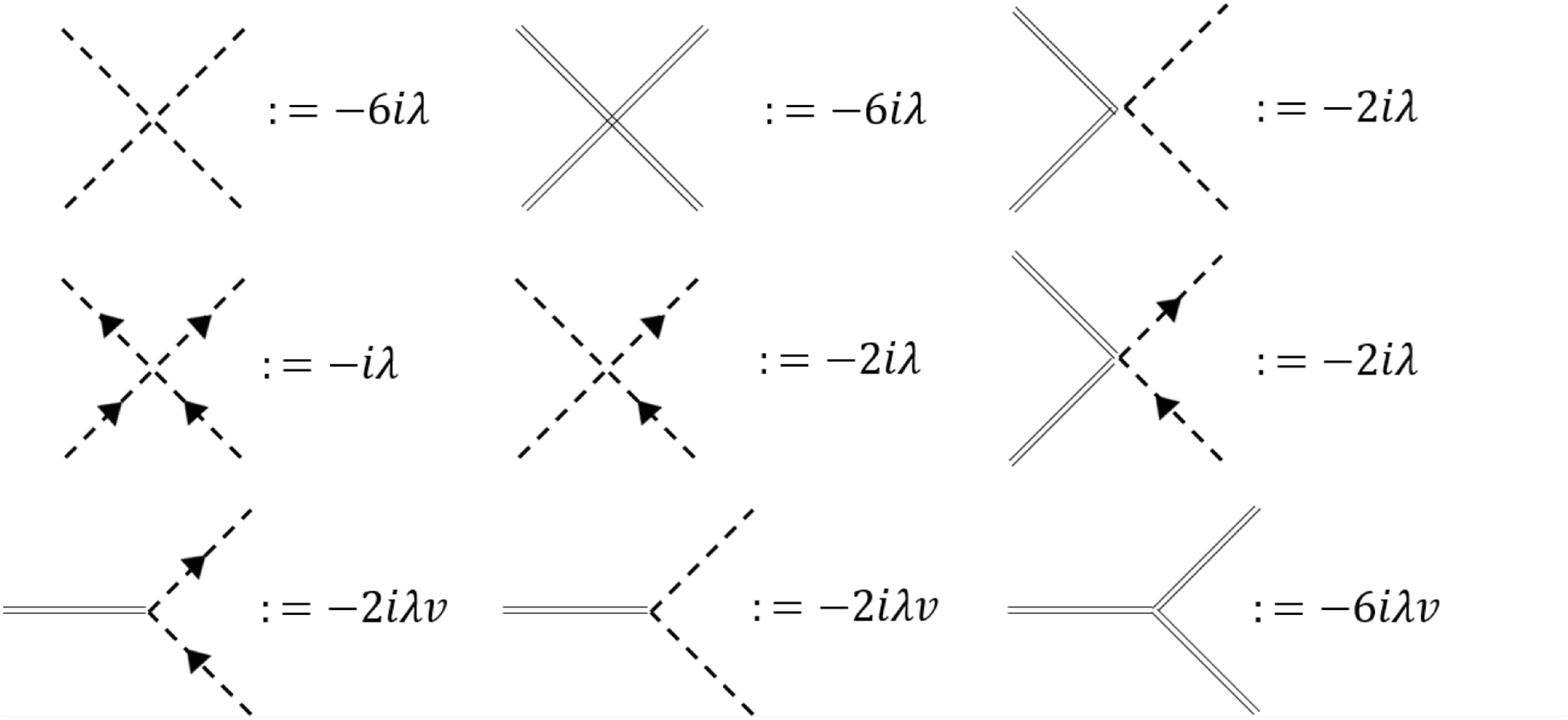}
    \caption{Meson interactions in the LSMq. Dashed lines are used to represent the neutral and charged pions  whereas double lines represent the $\sigma$.}
    \label{fig:mesonFeynmanrules}
\end{figure}
\begin{figure}[t]
    \includegraphics[scale=0.22]{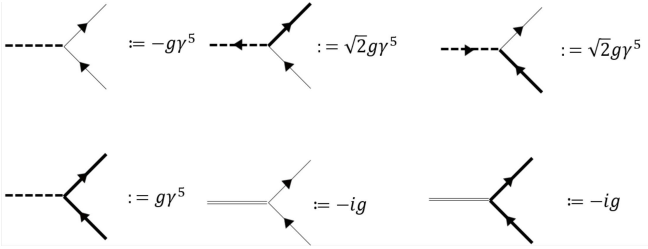}
    \caption{Quark-meson interactions in the LSMq. Dashed lines represent the neutral and charged pions whereas the double lines represent the $\sigma$. Solid lines represent the quarks. Thin solid lines represent the $d$ quark and thick solid lines the $u$ quark.}
    \label{fig:quarkFeynmanrules}
\end{figure}

We conclude this section by listing the Feynman rules deduced from the Lagrangian density in Eq.~\eqref{interactinglagrangian}. After accounting for the number of permutations for a set of equivalent lines and a factor of $i$ coming from the action, these are displayed in Fig.~\ref{fig:mesonFeynmanrules} and Fig.~\ref{fig:quarkFeynmanrules}. Fig.~\ref{fig:mesonFeynmanrules} shows the vertices arising in the meson sector and Fig.~\ref{fig:quarkFeynmanrules} shows the quark-meson vertices. Dashed lines represent the neutral and charged pions and double lines represent the $\sigma$ whereas thin solid lines represent the $d$ quark and thick solid lines the $u$ quark.
\section{\label{secIII} Magnetic field dependent boson and fermion propagators}
In order to consider the propagation of the charged modes within a magnetized background, we make the minimal substitution
\begin{eqnarray}
\partial_\mu\to D_\mu = \partial_\mu + iqA_\mu,
\label{minimal}    
\end{eqnarray}
where $q$ is the particle's electric charge and $A_\mu$ is the vector potential. Choosing the magnetic field to point in the direction of the $\hat{z}$-axis, namely $\vec{B}=B\hat{z}$, and working in an arbitrary gauge, we have
\begin{equation}
    A^{\mu}(x)=\frac{1}{2}x_{\nu}F^{\nu\mu}+\partial^{\mu}\Lambda(x),
    \label{arbitrarygauge}
\end{equation}
where $\Lambda$ is a well-behaved function which describes a gauge transformation from the symmetric gauge to an arbitrary gauge.

Notice that the ordinary derivative becomes the covariant derivative only for particles with non-vanishing electric charge. As a consequence, the propagation of charged bosons and fermions is described by propagators in the presence of a constant magnetic field. Using Schwinger's proper time representation, the fermion propagator can be written as 
\begin{equation}
    S(x,x')=e^{i\Phi(x,x')}S(x-x'),
    \label{fermionpropagatorincoordinatespace}
\end{equation}
where $\Phi(x,x')$ is the Schwinger's phase given by
\begin{eqnarray}
\Phi(x,x')=q\int_x^{x'}d\xi_\mu \left[
A^\mu(\xi) + \frac{1}{2}F^{\mu\nu}(\xi-x')_\nu
\right],
\label{phase}
\end{eqnarray}
and represents the translationally and gauge non-invariant part of the propagator in the presence of a magnetic background. Using Eq. \eqref{arbitrarygauge} into Eq. \eqref{phase}, the Schwinger's phase can be computed using the expression
\begin{equation}
    \Phi(x,x')=q\left[\frac{1}{2}x^{\mu}F_{\mu \nu }x'^{\nu}+\Lambda (x')-\Lambda(x) \right],
   \label{Schwingerphasegeneral}
\end{equation}
The translationally and gauge-invariant part of the propagator is provided by $S(x-x')$ that can be expressed in terms of its Fourier transform as
\begin{equation}
    S(x-x')=\int \frac{d^{4}p}{(2\pi)^{4}}S(p)e^{-ip\cdot(x-x')}, \label{Fouriertransformfermionpropagator}
\end{equation}
where
\begin{eqnarray}
    S(p)&=&\int_0^\infty \frac{ds}{\cos(|q_fB|s)}e^{is\left(p_\parallel^2-p_\perp^2\frac{\tan(|q_fB|s)}{|q_fB|s}-m_f^2+i\epsilon\right)}\nonumber\\
&\times&\left[
\Big(
\cos(|q_fB|s) + \gamma_1\gamma_2\sin(|q_fB|s)\text{sign}(q_fB)
\Big)\right.\nonumber\\
&\times&\left.\left(m_f +\slashed{p}_\parallel\right) - \frac{\slashed{p}_\perp}{\cos(|q_fB|s)}
\right].\label{fermionpropagatormomentumspace}
\end{eqnarray}
In a similar fashion,  for a charged scalar field we have  
\begin{eqnarray}
D(x,x')&=&e^{i\Phi(x,x')}D(x-x'),\nonumber\\
D(x-x')&=&\int \frac{d^{4}p}{(2\pi)^{4}}D(p)e^{-ip\cdot(x-x')},
\label{scalarprop}
\end{eqnarray}
with
\begin{eqnarray}
D(p)&=&\int_0^\infty \frac{ds}{\cos(|q_bB|s)}e^{is\left(p_\parallel^2-p_\perp^2\frac{\tan(|q_bB|s)}{|q_bB|s}-m_b^2+i\epsilon \right)},\nonumber\\
\label{bosonpropagatormomentumspace}
\end{eqnarray}
where the boson and fermion masses and electric charges are $m_b$, $q_b$ and $m_f$, $q_f$, respectively. 

The propagators in Eqs.~\eqref{fermionpropagatormomentumspace} and \eqref{bosonpropagatormomentumspace} can also be expanded as a sum over Landau levels. In this last representation, the expressions for the charged scalar and a fermion propagators are given by 
\begin{equation}
iD(p)=2i e^{-\frac{p_{\perp}^2}{|q_{b}B|}}\sum_{n=0}^{\infty}\frac{(-1)^{n}L_{n}^{0}\left(\frac{2p_{\perp}^{2}}{|q_{b}B|}\right)}{p_{\parallel}^2 -m_{b}^2 -(2n+1)|q_b B|+i\epsilon},
\label{bosonpropagatorLandauLevels}
\end{equation}
\begin{equation}
iS(p)=ie^{-\frac{p_{\perp}^2}{|q_{f}B|}}\sum_{n=0}^{\infty}\frac{(-1)^{n}D_{n}(p)}{p_{\parallel}^2 -m_{f}^2 -2n|q_f B|+i\epsilon},
\label{fermionpropagatorLandauLevels}
\end{equation}
where
\begin{eqnarray}
D_{n}(p)&=&2(\slashed{p}_\parallel +m_f)\mathcal{O}^{+} L_{n}^{0}\left(\frac{2p_{\perp}^2}{|q_{f}B|}\right)\nonumber\\
&-&2(\slashed{p}_\parallel +m_f)\mathcal{O}^{-} L_{n-1}^{0}\left(\frac{2p_{\perp}^2}{|q_{f}B|}\right)\nonumber\\
&+&4\slashed{p}_{\perp}L_{n-1}^{1}\left(\frac{2p_{\perp}^2}{|q_{f}B|}\right),
\label{fermionfact}
\end{eqnarray}
respectively, and $L_n^m(x)$ are the generalized Laguerre polynomials. In Eq.~(\ref{fermionfact}) the operators ${\mathcal{O}^{\pm}}$ are defined as
\begin{eqnarray}
{\mathcal{O}^{\pm}}=\frac{1}{2}\left( 1\pm i\gamma_1\gamma_2\, {\mbox{sign}}(qB)\right).
\label{Os}
\end{eqnarray}

We now proceed to use the interaction vertices and the magnetic field dependent  propagators to find the one-loop corrections to the boson self-coupling and boson-fermion coupling in the presence of a magnetic field.

\section{Magnetic corrections to the boson self-coupling}\label{secIV}
\begin{figure}[b]
    \includegraphics[scale=0.25]{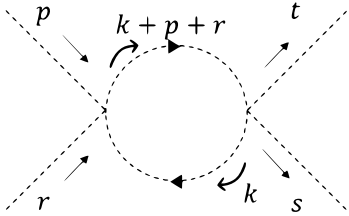}
    \caption{Feynman diagram representing the magnetic correction to the boson self-coupling at one loop. The loop particles are considered as electrically charged whereas the external ones can be either charged or neutral.}
    \label{fig:my_label}
\end{figure}
The magnetic field induced corrections to the boson self-coupling $\lambda$ can be obtained at one-loop order from the Feynman diagram depicted in Fig.~\ref{fig:my_label}, where the loop pions are the charged ones. In our approximation the external particles are taken as plane waves, that is, the states they represent do not experience the effects of the magnetic field. The only particles affected by the magnetic background are the charged, loop particles. With this approach we intend to capture the distinction between the modification of the interaction, that in a perturbative approach is a short distance effect, from the asymptotic propagation of the external particles, which corresponds to a long distance effect. Therefore, since the correction we look for is, in this sense, independent of whether the external bosons are charged or neutral, the electric charge of the external particles is irrelevant. Thus, the correction we look for is written as
\begin{eqnarray}
    -i6\lambda\Gamma_{\lambda}^{B}&=&\int \frac{d^{4}k}{(2\pi)^{4}}(-2i\lambda)iD_{\pi^{-}}(k)(-2i\lambda)\nonumber\\
    &\times& iD_{\pi^{-}}(k+p+r)+{\mbox{CC}},
    \label{magneticlambda}
\end{eqnarray}
where CC denotes the charge conjugate term and the subindex in the boson propagator indicates the propagating species. Notice that since the loop involves the same propagating particle, the Schwinger's phase vanishes.

According to the explicit computation shown in Appendix \ref{computationmagneticlambda} and in the {\it static} limit $p_{0},r_{0}\rightarrow 0$, $\vec{p}=\vec{r}=\vec{0}$, we obtain
\begin{eqnarray}
    \Gamma_{\lambda}^{B}&=&-\frac{\lambda}{12\pi^{2}} \bigg[\frac{1}{\varepsilon}-\gamma_{E}+\ln{(4\pi)}-\psi^{0}\left(\frac{|q_{b}B|+m_{\pi}^{2}}{2|q_{b}B|}\right) \nonumber\\
    &+&\ln{\left(\frac{\mu^{2}}{2|q_{b}B|} \right)}\bigg],
    \label{finalresultlambda}
\end{eqnarray}
where $\psi^{0}$ is the digamma function and $|q_{b}B|=|eB|$. In the modified minimal subtraction scheme $\overline{MS}$, the first three terms in Eq.~\eqref{finalresultlambda} are associated to the corresponding vertex counter-term. Therefore, the finite magnetic correction to the boson self-coupling is given by
\begin{equation}
    \Gamma_{\lambda}^{B}=-\frac{\lambda}{12\pi^{2}} \bigg[\ln{\left(\frac{\mu^{2}}{2|q_{b}B|} \right)}-\psi^{0}\left(\frac{|q_{b}B|+m_{\pi}^{2}}{2|q_{b}B|}\right)\bigg].
    \label{finiteresultlambda}
\end{equation}
Notice that the result in Eq.~\eqref{finiteresultlambda} depends on the ultraviolet renormalization scale $\mu$. In order to gain some insight on the appropriate choice of this scale, we can compare this result with the one obtained in the strong field limit where, as a good approximation, one can consider just the lowest Landau level (LLL) contribution, $n=0$, for the charged boson propagators of Eq.~\eqref{bosonpropagatorLandauLevels}, namely
\begin{equation}
    iD^{LLL}(p)=\frac{2ie^{-\frac{p_{\perp}^{2}}{|q_{b}B|}}}{p_{\parallel}^{2}-m_{b}^{2}-|q_{b}B|+i\epsilon}.
    \label{LLLbos}
\end{equation}
Therefore, using Eq.~(\ref{LLLbos}) into Eq.~\eqref{magneticlambda}, and working also in the static limit, the magnetic correction to the boson self-coupling in the LLL is given by 
\begin{equation}
    \Gamma_{\lambda}^{LLL}=-\frac{\lambda}{6\pi^{2}}  \frac{|q_{b}B|}{|q_{b}B|+m_{\pi}^{2}},
    \label{magneticcorrectiontolambdaLLLallmomentazero}
\end{equation}
which is independent of $\mu$. On the other hand, in the absence of a magnetic field, it is easy to show that the one-loop correction to the boson self-coupling is given by
\begin{equation}
    \Gamma_{\lambda}=-\frac{\lambda}{12\pi^{2}}\ln{\left(\frac{\mu^{2}}{m_\pi^2} \right)}.
    \label{vaconeloop}
\end{equation}
Notice that in order to obtain the limits when $|q_bB|\to 0$ in Eq.~(\ref{vaconeloop}), and when  $|q_bB|\to\infty$ in Eq.~(\ref{magneticcorrectiontolambdaLLLallmomentazero}), from the arbitrary field strength result of Eq.~(\ref{finiteresultlambda}), it is necessary that $\mu$ depends on $|q_bB|$. In fact, the match is obtained when $\mu^2$ is explicitly chosen as $\mu^{2}=m_\pi^2+2|q_{b}B|$, for which the arbitrary field strength result becomes
\begin{eqnarray}
    \Gamma_{\lambda}^{B}=-\frac{\lambda}{12\pi^{2}} \bigg[\ln{\left(\frac{m_\pi^{2}+2|q_bB|}{2|q_bB|} \right)}-\psi^{0}\left(\frac{|q_bB|+m_{\pi}^{2}}{2|q_bB|}\right)\bigg].\nonumber\\
    \label{exactmuexpl}
\end{eqnarray}
\begin{figure}[t]
    \includegraphics[scale=0.85]{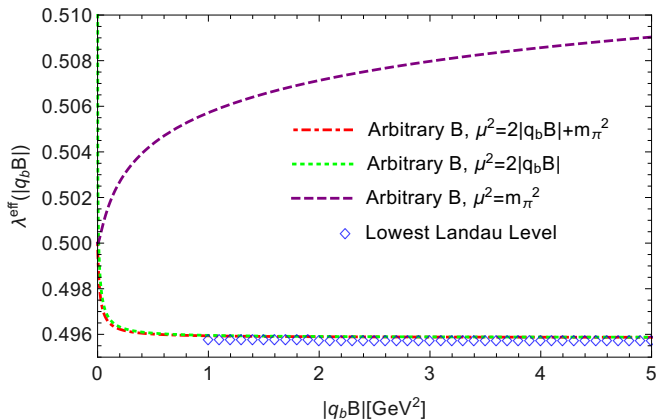}
    \caption{Comparison of the magnetic field dependence of the effective boson self-coupling $\lambda^{eff}=\lambda(1+\Gamma_\lambda^B)$ in the arbitrary field approach and the strong field limit, both computed in the static limit. For the calculation we use $\lambda=0.5$ and $m_{\pi}=0.140$ GeV. Shown are the cases where for the arbitrary field intensity calculation, the ultraviolet renormalization scale $\mu^2$ is taken as $m_\pi^2+2|q_bB|$ (red dashed line),  $\mu^{2}=2|q_bB|$ (green dotted line) and a fixed value $\mu^{2}=m_\pi^2$ (purple dashed line). Notice that, although the choice $\mu^{2}=2|q_bB|$ does a good description of the LLL result for large field strengths, when $|q_bB|=0$ the effective coupling diverges which signals that this choice is not appropriate. For the rest of the cases, the self-coupling relative change from the vacuum value is rather small, of order 0.8\%.}
    \label{fig:lambdacomparison}
\end{figure}
With this choice, the result reproduces the behavior of the coupling in both extreme limiting values of $|q_bB|$ and it is also compatible with the behavior of the coupling found in Ref.~\cite{pionmassmag} for the weak field case. This behavior is shown in Fig.~\ref{fig:lambdacomparison}  where we plot the effective, magnetic field dependent boson self-coupling $\lambda^{eff}=\lambda\left(1 + \Gamma_\lambda^B\right)$ as a function of the field strength. In contrast, when $\mu$ is taken at a fixed value, the arbitrary field result does not match the LLL case. We interpret this result as signaling that when the field strength is the largest energy scale, $\mu$ needs to be taken also as this large scale since otherwise the computation is not consistent when the strength of the magnetic field surpasses a given fixed scale. At the same time, when the field strength vanishes, the only remaining energy scale is the pion mass and $\mu$ needs to be taken solely as this energy scale. Furthermore, notice that $2|q_bB|$ corresponds to the square of the energy gap between Landau levels and thus that in order for $\mu$ to correspond to the largest energy scale, it is important that for large values of the field strength, $\mu^2$ is taken as the square of this energy gap. In contrast, as also shown in Fig.~\ref{fig:lambdacomparison}, the usual prescription~\cite{scoccola01,Scoccola}, whereby one just subtracts the vacuum correction, represented by the purple dashed line computed with $\mu^2=m_\pi^2$, behaves  opposite  to what is expected from the result obtained using the LLL propagator. Since the latter provides a reliable approximation for large field strengths, we conclude that a simple vacuum subtraction prescription leads to a non-reliable limit for large values of the field strength.

\section{Magnetic corrections to the boson-fermion coupling}\label{secV}
\begin{figure}[b]
\includegraphics[scale=0.3]{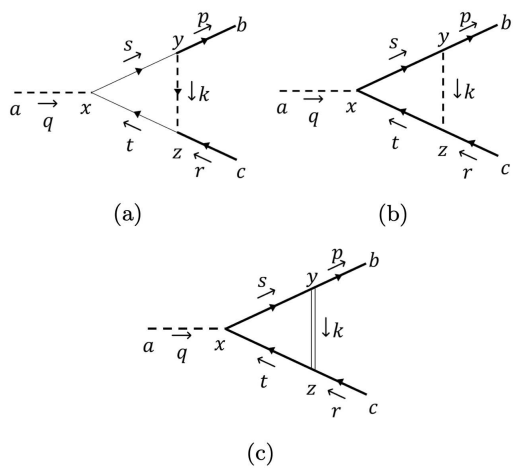}
\caption{Feynman diagrams that contribute to the boson-fermion coupling at one-loop order. The diagrams show the case with a neutral pion and two $u$ quarks as the external particles.}
    \label{fig:Feynmandiagramsg}
\end{figure}

The magnetic corrections to the coupling constant $g$ at one-loop level can be obtained from the sum of the three Feynman diagrams depicted in Fig.~\ref{fig:Feynmandiagramsg}.
Since the correction can be obtained from the sum of the allowed Feynman diagrams coupling one boson and two quarks, here we consider the magnetic correction to the boson-fermion coupling for the choice of external particles shown in Fig.~\ref{fig:Feynmandiagramsg}. Also since, as discussed in the previous section, the use of propagators in the LLL approximation provide a reliable description in the case of the strong field limit, we hereby restrict ourselves to this case using the LLL approximation, Eq.~(\ref{LLLbos}), for the boson propagator and the fermion propagator also in the LLL, given by
\begin{equation}
    iS^{LLL}(k)=2ie^{\frac{-k_{\perp}^{2}}{|q_{f}B|}}\frac{\slashed{k}_{\parallel}+m_{f}}{k_{\parallel}^{2}-m_{f}^{2}+i\epsilon}\mathcal{O}^{\pm}.
    \label{fermionpropagatorLLL}
\end{equation}

We start by computing the contribution from the diagram in Fig.~\ref{fig:Feynmandiagramsg}(a). We first compute the quantity $I^{B}_{1,g}$ which is given explicitly by 
\begin{equation}
\begin{split}
I_{1,g}^{B}&=\int d^{4}x\;d^{4}y\;d^{4}x \int \frac{d^{4}s}{(2\pi)^{4}}\frac{d^{4}t}{(2\pi)^{4}}\frac{d^{4}k}{(2\pi)^{4}}\;e^{i\Phi_{1,l}} e^{-ip\cdot y}\\
&\times\left(\sqrt{2}g\gamma^{5} \right)e^{-is\cdot(x-y)}iS_{d}(s)\left(-g\gamma^{5}\right) e^{iq\cdot x} e^{-it\cdot(z-x)}\\
&\times iS_{d}(t)\left(\sqrt{2}g\gamma^{5} \right) e^{-ik\cdot(y-z)}iD_{\pi^{-}}(k)e^{ir\cdot z}+{\mbox{CC}}.
\label{Ig1}   
\end{split}
\end{equation}
The information from the Schwinger phases is contained in the function $\Phi_{1,l}(x,y,z)$. This function depends on the space-time points located at the vertices. For the calculation to have a solid physical meaning, this phase should be a gauge invariant quantity. We proceed to show this fact explicitly. 

Notice that the {\it total} Schwinger phase $\Phi_{1,t}$ associated to the Feynman diagram in Fig.~\ref{fig:Feynmandiagramsg}(a), contains not only the information of the space-time points at the interaction vertices $x,y,z$, but also the one coming from the {\it external} space-time points $a,b,c$. Therefore $\Phi_{1,t}$ is given explicitly by
\begin{eqnarray}
\Phi_{1,t}&=& \Phi_{d}(x,y)+\Phi_{\pi^{-}}(y,z)+\Phi_{d}(z,x)+\Phi_{u}(y,b)\nonumber\\
&+&\Phi_{u}(c,z).
\label{Phi1t}
\end{eqnarray}
Using Eq.~\eqref{Schwingerphasegeneral} into Eq.~\eqref{Phi1t} we have
\begin{eqnarray}
\Phi_{1,t}&=&-\frac{1}{2}q_{d}F_{\mu \nu }\left(y^{\mu}x^{\nu}+x^{\mu}z^{\nu}\right) -\frac{1}{2}q_{\pi^{-}}F_{\mu \nu }z^{\mu}y^{\nu} \nonumber\\
&-&\frac{1}{2}q_{u}F_{\mu \nu }\left(b^{\mu}y^{\nu}+z^{\mu}c^{\nu} \right)+q_{u}[\Lambda(b)-\Lambda (c)].
    \label{Schwingerphasetotal1}
\end{eqnarray}
Notice that terms depending on $\Lambda$ evaluated at the internal space-time points add up to zero. Therefore, the integration over the configuration space becomes independent of the gauge choice. However, this would not be the case were we just to consider the phase factors associated to the particles within the loop, since the result of the integration would then become gauge dependent. This  observation is essential since otherwise one faces a non-conservation of electric charge at each vertex when just considering the phases within the loop. On the other hand, Eq.~\eqref{Schwingerphasetotal1} contains a mixing between the phases associated to loop particles, $\Phi_{1,l}$, and the phases from  external particles, $\Phi_{ext}$, where the last term is associated to the external charged lines in the diagram and can be written as 
\begin{equation}
  \Phi_{ext}=-\frac{1}{2}q_{u}F_{\mu \nu}\left(b^{\mu}x^{\nu}+x^{\mu}c^{\nu} \right)+q_{u}[\Lambda(b)-\Lambda(c)].
  \label{externalSchwingerphase}
\end{equation} 
In order to separate these contributions we write
\begin{equation}
   \Phi_{1,t}=\Phi_{1,l}+\Phi_{ext},
\end{equation}
we resort to consider that the external particles can be described as plane waves. Physically, this means that we consider the propagation of the external particles during short distances and times. In this manner we neglect long distance effect introduced when the magnetic field acts over the external particles. Therefore, we can take $y^{\mu}\approx b^{\mu}$ and $z^{\mu}\approx c^{\mu}$ such that
\begin{eqnarray}
\Phi_{1,t}&=&-\frac{1}{2}q_{u}F_{\mu \nu}\left(b^{\mu}x^{\nu}+x^{\mu}c^{\nu} \right)+q_{u}[\Lambda(b)-\Lambda(c)],\nonumber\\
&-&\frac{1}{2}q_{\pi^{-}}F_{\mu\nu}\left(y^{\mu}x^{\nu}+z^{\mu}y^{\nu}+x^{\mu}z^{\nu} \right).\label{Schwingerphasetotal1withapproximation}  
\end{eqnarray}
Using this approximation we can separate the phase factors coming from external and internal, loop particles. Thus, for the computation of the magnetic field correction for the coupling $g$, we need only to account for the last term in Eq.~\eqref{Schwingerphasetotal1withapproximation} whereas the first and second terms in Eq.~\eqref{Schwingerphasetotal1withapproximation} are associated to the external phase given by Eq.~\eqref{externalSchwingerphase}. Therefore we have
\begin{equation}
    \Phi_{1,l}=-\frac{1}{2}q_{\pi^{-}}F_{\mu\nu}\left(y^{\mu}x^{\nu}+z^{\mu}y^{\nu}+x^{\mu}z^{\nu} \right).
\end{equation}
It is important to note that the contribution from the Schwinger phase is gauge-invariant.
Using that $F_{21}=-F_{12}=|B|$ and $q_{\pi^{-}}=-|e|$, we get
\begin{equation}
    \Phi_{1,l}=\frac{1}{2}|eB|\varepsilon_{ij}\left(x_{i}y_{j}+y_{i}z_{j}+z_{i}x_{j} \right),\quad i,j=1,2,
\end{equation}
where $\varepsilon_{ij}$ is the Levi-Civita symbol.
Having identified the Schwinger's phase contribution, we can perform the integration over coordinates. On doing so, we obtain the energy-momentum conservation for the external particles, and can write 
\begin{equation}
    I_{1,g}^{B}=(2\pi)^{4}\delta^{(4)}(p-r-q)g\gamma^{5}\Gamma_{1,g}^{B},
\end{equation}
where $g\gamma^{5}\Gamma_{1,g}$ is identified as the contribution to the magnetic field correction to the vertex, given explicitly by
\begin{eqnarray}
   g\gamma^{5}\Gamma_{1,g}^{B}&=&\int \frac{d^{2}s_{\perp} d^{2}t_{\perp}}{\pi^{2}|eB|^2} \frac{d^{4}k}{(2\pi)^{4}}\left(\sqrt{2}g\gamma^{5} \right)iS_{d}(k_{\parallel}+p_{\parallel},s_{\perp}) \nonumber\\
   &\times&\left(-g\gamma^{5}\right) iS_{d}(k_{\parallel}+r_{\parallel},t_{\perp})\left(\sqrt{2}g\gamma^{5} \right)iD_{\pi^{-}}(k_{\parallel},k_{\perp})\nonumber\\ 
&\times&  e^{i\frac{2}{|eB|}\varepsilon_{ij}(s-q-t)_{i}(s-p-k)_{j}}+ {\mbox{CC}}. 
\label{Gammag1}
\end{eqnarray}
Following the procedure explicitly shown in Appendix \ref{computationmagneticg} and the static limit $p_{0}=r_{0}=m_{f}$ and $\vec{p}=\vec{r}=\vec{0}$ we get
\begin{eqnarray}
    \Gamma_{1,g}^{LLL}&=&\frac{g^{2}|eB|}{16\pi^{2}m_{f}^{2}}\int_{0}^{1}du\frac{u}{ u^{2}+\alpha(1-u)}\nonumber\\
    &\times&\left[1+\frac{(2-u)u}{u^{2}+\alpha(1-u)} \right],
    \label{resultGammag1}
\end{eqnarray}
where $\alpha=(m_{\pi}^{2}+|eB|)/m_{f}^{2}$.

Next, we compute the contribution from the Feynman diagram depicted in Fig.~\ref{fig:Feynmandiagramsg}(b). This contribution can be obtained from the function $I_{2,g}^{B}$, which is given by 
\begin{equation}
\begin{split}
I_{2,g}^{B}&=\int d^{4}x\;d^{4}y\;d^{4}z \int \frac{d^{4}s}{(2\pi)^{4}}\frac{d^{4}t}{(2\pi)^{4}}\frac{d^{4}k}{(2\pi)^{4}}e^{i\Phi_{2,l}} e^{-ip\cdot y}\\
&\times\left(g\gamma^{5} \right)e^{-is\cdot(x-y)}iS_{u}(s)\left(g\gamma^{5}\right)e^{iq\cdot x} e^{-it\cdot(z-x)}\\ 
& \times   iS_{u}(t)\left(g\gamma^{5} \right) e^{-ik\cdot(y-z)}iD_{\pi^{0}}(k)e^{ir\cdot z}+{\mbox{CC}}.
\label{Ig2}
\end{split}
\end{equation}
In a similar fashion, we first compute the Schwinger's phase associated to the whole diagram in Fig.~\ref{fig:Feynmandiagramsg}(b), namely,
\begin{equation}
    \Phi_{2,t}= \Phi_{u}(x,y)+\Phi_{u}(z,x)+\Phi_{u}(y,b)+\Phi_{u}(c,z).
    \label{Schwingerphasetriangle2}
\end{equation}
Using Eq.~\eqref{Schwingerphasegeneral} into Eq.~\eqref{Schwingerphasetriangle2}, we get
\begin{eqnarray}
\Phi_{2,t}&=&-\frac{1}{2}F_{\mu \nu}q_{u}\left(y^{\mu}x^{\nu}+b^{\mu}y^{\nu}+x^{\mu}z^{\nu}+z^{\mu}c^{\nu}\right)\nonumber\\
&+&q_{u}\left[\Lambda(b)-\Lambda(c)\right].
\label{Schwingerphasetotal2}   
\end{eqnarray}
Once again terms that depend on  $\Lambda$, evaluated at internal points, vanish. On the other hand, the Schwinger's phase associated to the tree level diagram is given by Eq.~\eqref{externalSchwingerphase}. Adding and subtracting the first term from this equation to Eq.~\eqref{Schwingerphasetotal2} we have
\begin{eqnarray}
\Phi_{2,t}&=&-\frac{1}{2}F_{\mu \nu}q_{u}\left(y^{\mu}x^{\nu}+b^{\mu}y^{\nu}+x^{\mu}z^{\nu}+z^{\mu}c^{\nu}\right)\nonumber \\
&+&q_{u}\left[\Lambda(b)-\Lambda(c)\right]-\frac{1}{2}q_{u}F_{\mu\nu}\left(b^{\mu}x^{\nu}+x^{\mu}c^{\nu} \right)\nonumber\\
&+&\frac{1}{2}q_{u}F_{\mu\nu}\left(b^{\mu}x^{\nu}+x^{\mu}c^{\nu} \right).    
\end{eqnarray}
Assuming that $y^{\mu}\approx b^{\mu}$ and $z^{\mu}\approx c^{\mu}$ (short space-time interval propagation after the interaction) we can write
\begin{equation}
   \Phi_{2,t}=-\frac{1}{2}q_{u}F_{\mu\nu}\left(b^{\mu}x^{\nu}+x^{\mu}c^{\nu} \right)+q_{u}\left[\Lambda(b)-\Lambda(c)\right].
\end{equation}
This result coincides with Eq. \eqref{externalSchwingerphase}. Therefore, we can conclude that the Schwinger's phase associated to the loop particles vanishes
\begin{equation}
     \Phi_{2,l}=0.
\end{equation}
Upon integration over configuration space, we can identify the contribution to the magnetic correction from this diagram, $g\gamma^{5}\Gamma_{2,g}$, as
\begin{equation}
    I_{2,g}^{B}=(2\pi)^{4}\delta^{(4)}(p-r-q)g\gamma^{5}\Gamma_{2,g}.  
\end{equation}
Again, notice that using this approximation we recover the energy-momentum conservation for the external particles, whereas the magnetic correction is associated to the loop and can be expressed as
\begin{eqnarray}
    g\gamma^{5}\Gamma_{2,g}^{B}&=&\int\frac{d^{4}k}{(2\pi)^{4}} \left(g\gamma^{5} \right)iS_{u}(k+p)\left(g\gamma^{5}\right)iS_{u}(k+r) \nonumber\\
    &\times&\left(g\gamma^{5} \right) iD_{\pi^{0}}(k)+{\mbox{CC}}.
    \label{Gammag2}
\end{eqnarray}
The computation of this quantity is explicitly performed in Appendix \ref{computationmagneticg} in the strong field limit and can be expressed as
\begin{eqnarray}
   \Gamma_{2,g}^{LLL}&=&-\frac{g^{2}}{2\pi^{2} m_{f}^{2}}\int_{0}^{1}du \int_{0}^{\infty} dk_{\perp}\; k_{\perp} e^{-\frac{3k_{\perp}^{2}}{|eB|}} \nonumber\\
   &\times& \frac{u}{u^{2}+\beta(1-u)}\left[1+\frac{(2-u)u}{ u^{2}+\beta(1-u)} \right],
   \label{resultGammag2}
\end{eqnarray}
where $\beta= (k_{\perp}^{2}+m_{\pi}^{2})/m_{f}^{2}$. 

The diagram in Fig.~\ref{fig:Feynmandiagramsg}(c) can be computed from the quantity $I_{3,g}^{B}$, given explicitly by 
\begin{eqnarray}
I_{3,g}^{B}&=&\int d^{4}x\;d^{4}y\;d^{4}z \int \frac{d^{4}s}{(2\pi)^{4}}\frac{d^{4}t}{(2\pi)^{4}}\frac{d^{4}k}{(2\pi)^{4}}e^{i\Phi_{3,l}} e^{-ip\cdot y} \nonumber \\
&\times&\left(-ig \right)e^{-is\cdot(x-y)}iS_{u}(s)\left(g\gamma^{5}\right) e^{iq\cdot x} e^{-it\cdot(z-x)} \nonumber\\
&\times& iS_{u}(t)\left(-ig \right) e^{-ik\cdot(y-z)}iD_{\sigma}(k)e^{ir\cdot z}+{\mbox{CC}}.
\label{Ig3}
\end{eqnarray}
In a similar fashion, one can compute the Schwinger's phase from this loop, $\Phi_{3,l}(x,y,z)$. It is easy to see that this phase satisfies $\Phi_{3,t}=\Phi_{2,t}$ and therefore, the internal Schwinger's phase vanishes when considering short-range propagation of the external particles, namely,
\begin{equation}
     \Phi_{3,l}=0.
\end{equation}
After performing the integration over the configuration space, we obtain a the relation between $I_{3,g}^{B}$ and the contribution to the magnetic correction to the boson-fermion coupling, $g\gamma^{5}\Gamma_{3,g}$, given by
\begin{equation}
    I_{3,g}^{B}=(2\pi)^{4}\delta^{(4)}(p-r-q)g\gamma^{5}\Gamma_{3,g}, 
\end{equation}
with
\begin{eqnarray}
g\gamma^{5}\Gamma_{3,g}^{B}&=& \int \frac{d^{4}k}{(2\pi)^{4}} \left(-ig \right)iS_{u}(k+p)\left(g\gamma^{5}\right)iS_{u}(k+r)\nonumber\\
&\times&\left(-ig \right) iD_{\sigma}(k)+{\mbox{CC}}.
\end{eqnarray}
Once again, using the LLL propagators and following the explicit procedure shown in Appendix \ref{computationmagneticg} we get
\begin{eqnarray}
   \Gamma_{3,g}^{LLL}&=&\frac{g^{2}}{2\pi^{2}m_{f}^{2}}\int_{0}^{1}du \int_{0}^{\infty} dk_{\perp}\;
    k_{\perp}e^{-\frac{3k_{\perp}^{2}}{|eB|}} \nonumber\\
    &\times&\frac{u}{u^{2}+\gamma(1-u)}\left[1+\frac{(2-u)u}{u^{2}+\gamma(1-u)} \right],
    \label{resultGammag3}   
\end{eqnarray}
where $\gamma=(k_{\perp}^{2}+m_{\sigma}^{2})/m_{f}^{2}$.

\begin{figure}[t]
    \includegraphics[scale=0.85]{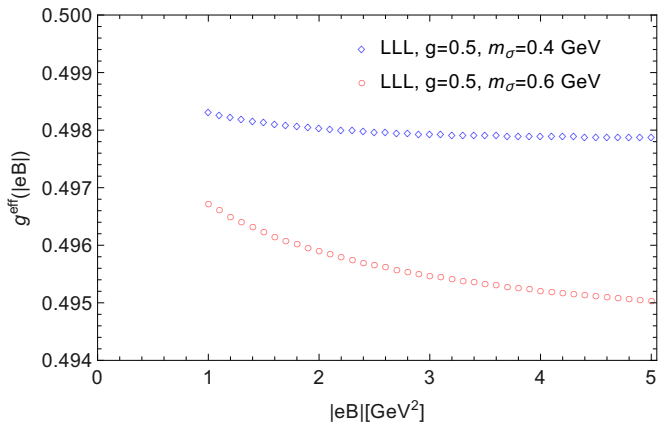}
    \caption{Magnetic field dependence of the effective boson-fermion coupling $g^{eff}=g(1+\Gamma_{g}^B)$ in in the static limit and the strong field approximation. For the calculation we used $g=0.5$, $m_{\pi}=0.140$ GeV, $m_{f}=0.3$ GeV and the two values $m_{\sigma}=0.4,\,0.6$ GeV. In both cases $g^{eff}$ monotonically decreases in an interval $|eB| = 1-5$ GeV$^{2}$. Notice that the relative change with regard to the vacuum is of order $0.5\%$ and $1\%$, respectively.}
    \label{fig:gcomparison}
\end{figure}

The total magnetic correction to the boson-fermion coupling in the strong field limit is given by the sum of the three contributions, namely,
\begin{equation}
    \Gamma_{g}^{LLL}=\Gamma^{LLL}_{1,g}+\Gamma^{LLL}_{2,g}+\Gamma^{LLL}_{3,g}.
\end{equation}
The effective boson-fermion coupling, $g^{eff}$ is thus given by
\begin{equation}
    g^{eff}=g\left(1+\Gamma_{g}^{LLL} \right).
\end{equation}
Figure~\ref{fig:gcomparison} shows the behavior of the boson-fermion coupling as a function of the field strength. For the calculation we set $m_{\pi}=0.140$ GeV, $m_{f}=0.3$ GeV and $m_{\sigma}=0.4,\,0.6$ GeV. Notice that the coupling decreases monotonically over a large range of the field strength. However, the relative change is rather small.

\section{\label{concl}Conclusions and outlook}

In this work, we have computed the magnetic field induced corrections to the boson self-coupling and to the boson-fermion coupling in the LSMq, in the static limit. For the former, we have performed the computation for an arbitrary field strength as well as in the strong field approximation. For the latter we worked in the strong field limit. We have shown that the full magnetic field corrections for the boson self-coupling depends on the ultraviolet renormalization scale and that this dependence cannot be removed by the usual vacuum subtraction. The reason for this behavior is that for a fixed ultraviolet renormalization scale, the calculation is not valid any longer when the field strength surpasses that fixed value, thus becoming the largest energy scale. Taking as a guide the result in the strong field limit, we have found the appropriate choice for the renormalization scale that produces the expected behavior for the two extreme limits, namely, when the field vanishes or when this becomes very large. 

For the calculation of the effective boson-fermion coupling, we have shown that when considering that the external, charged particles, propagate only during short space-time intervals, the effects coming from the Schwinger's phase become gauge invariant and that the usual energy-momentum conservation can be factored out from the vertex function. 

Recall that vertex corrections are functions of arbitrary values of the external particle's momenta. In this work we have considered the static limit approximation whereby the vertices are computed for vanishing external momenta. We emphasize that this approximation goes in hand with the short distance approximation whereby the motion of these particles is considered to happen during very short times so as to also neglect the magnetic field effects that would otherwise build up during large times, deviating the motion of external charged particles from free propagation. Within these approximations the effective boson self-coupling and boson-fermion coupling show a modest monotonic decrease over a large interval of magnetic field strengths.

The results of this work can now be used to find the corrections to the mass of neutral and charged pions introduced by magnetic field effects. Other possible scenarios of physical interest where the findings of this work can have a potential impact include the properties of the nuclear equation of state within dense and compact astrophysical objects, such as the core of neutron stars, which is affected by magnetic field dependent baryon and mesons masses and couplings, and the shear and bulk viscosity in quark-meson matter. The first of these topics is currently being actively pursued and will be soon reported elsewhere.

\section*{Acknowledgements}
The authors are in debt with Jorge Casta\~no for useful discussions. This work was supported in part by UNAM-DGAPA-PAPIIT grant number IG100219. L. A. H. acknowledges support from a PAPIIT-DGAPA-UNAM fellowship. R. Zamora acknowledges support from FONDECYT (Chile) under grant No. 1200483. This work is partially supported by Conselho Nacional de Desenvolvimento Cient\'ifico e Tecnol\'ogico  (CNPq), Grant No. 304758/2017-5 (R. L. S. F); Funda\c{c}\~ao de Amparo \`a Pesquisa do Estado do Rio Grande do Sul (FAPERGS), Grants Nos. 19/2551- 0000690-0 and 19/2551-0001948-3 (R. L. S. F.).

\begin{appendix}
\section{Magnetic corrections to the boson self-coupling}\label{computationmagneticlambda}
To compute the magnetic correction to the boson self-coupling, we start from the Landau level representation of the charged boson propagator in Eq.~\eqref{bosonpropagatorLandauLevels} and use it into the expression for the magnetic correction to $\lambda$ given by
\begin{eqnarray}
    -i6\lambda\Gamma_{\lambda}^{B}&=&\int \frac{d^{4}k}{(2\pi)^{4}}(-2i\lambda)iD_{\pi^{-}}^{B}(k)(-2i\lambda)\nonumber\\
    &\times& iD_{\pi^{-}}^{B}(k+p+r)+{\mbox{CC}}.
    \label{appmagneticlambda}
\end{eqnarray}
Performing a Wick rotation in $k$ and $s=r+p$, such that $k_{0}\rightarrow ik_{4}$ and $s_{0} \rightarrow is_{4}$, then 
\begin{equation}
    k_{\parallel}^{2}\rightarrow-k_{E\parallel}^{2}, \; (k+s)_{\parallel}^{2}\rightarrow -(k+s)_{E\parallel}^{2},\; d^{4}k\rightarrow id^{4}k_{E}.
\end{equation}
We now introduce two Schwinger parameters, $x_{1}, \ x_{2}$, $d^{2}x=dx_{1}dx_{2}$ such that the magnetic correction can be written as
\begin{equation}
\begin{split}
    \Gamma_{\lambda}^{B}&=-\frac{16}{3}\lambda\int\frac{d^{4}k_{E}}{(2\pi)^{4}}\int d^{2}x \sum_{n,m=0}^{\infty}r_{1}^{n}r_{2}^{m}L_{n}^{0}\left(s_{1}\right)L_{m}^{0}\left(s_{2}\right)\\ &\times e^{-\frac{k_{\perp}^{2}}{|q_{b}B|}-\frac{(k+s)_{\perp}^{2}}{|q_{b}B|}-x_{1}[\alpha(k_{E\parallel})+|q_{b}B|]-x_{2}[\beta(k_{E\parallel})+|q_{b}B|]},
\end{split}
\end{equation}
where
\begin{eqnarray}
    &&s_{1}= 2k_{\perp}^{2}/|q_{b}B|,\; s_{2}= 2(k+s)_{\perp}^{2}/|q_{b}B|,\nonumber\\
    &&\alpha(k_{E\parallel})=k_{E\parallel}^{2}+m_{\pi}^{2}-i\epsilon,\;\nonumber\\
    &&\beta(k_{E\parallel})=(k+s)_{E\parallel}^{2}+m_{\pi}^{2}-i\epsilon,
\end{eqnarray}
and $r_{i}=-e^{-2|q_{b}B|x_{i}},$ $i=1,2$. Using the generating function of Laguerre polynomials
\begin{equation}
    \sum_{n=0}^{\infty}r_{i}^{n}L_{n}^{0}(s_{i})=\frac{1}{1-r_{i}}e^{-\frac{r_{i}}{1-r_{i}}s_{i}},
\end{equation}
we obtain 
\begin{equation}
    \Gamma_{\lambda}^{B}=-\frac{16}{3}\lambda \int d^{2}x \frac{e^{-(x_{1}+x_{2})|q_{b}B|}}{(1-r_{1})(1-r_{2})} I(x_{1},x_{2})J(x_{1},x_{2}),
        \label{lambda2}
\end{equation}
where we define 
\begin{eqnarray}
    &&I(x_{1},x_{2})=\int\frac{d^{2}k_{\perp}}{(2\pi)^{2}}e^{-\frac{k_{\perp}^{2}}{|q_{b}B|}(1-2\eta(x_{1}))-\frac{(k+s)_{\perp}^{2}}{|q_{b}B|}(1-2\eta(x_{2}))},\nonumber\\
    &&J(x_{1},x_{2})= \mu^{4-d}\int\frac{d^{d-2}k_{E\parallel}}{(2\pi)^{d-2}}e^{-x_{1}\alpha(k_{E\parallel})-x_{2}\beta(k_{E\parallel})},\nonumber\\
    &&\eta(x_{i})= \frac{1}{e^{2|q_{b}B|x_{i}}+1},
\end{eqnarray}
with $i=1,2$ and $\epsilon \rightarrow 0$. To carry out the integrals, we use  dimensional regularization, namely
\begin{equation}
    \int\frac{d^{4}k_{E}}{(2\pi)^{4}}\rightarrow \mu^{4-d}\int\frac{d^{d-2}k_{E\parallel}}{(2\pi)^{d-2}}\int\frac{d^{2}k_{\perp}}{(2\pi)^{2}}.
\end{equation}
First, to find $I(x_{1},x_{2})$ we consider the change of variable
\begin{equation}
    q_{\perp}=k_{\perp}+\frac{1-2\eta(x_{2})}{2(1-\eta(x_{1})-\eta(x_{2}))}s_{\perp},
\end{equation}
and the identity
\begin{equation}
    1-2\eta(x_{i})=\tanh(|q_{b}B|x_{i}).
\end{equation}
Completing the square we have
\begin{eqnarray}
    \!\!\!\!\!\!\!\!&&I(x_{1},x_{2})=\frac{|q_{b}B|/4\pi}{\tanh(|q_{b}B|x_{1})+\tanh(|q_{b}B|x_{2})}\nonumber\\
    \!\!\!\!\!\!\!\!&\times&\exp\left[-\frac{\tanh(|q_{b}B|x_{1})\tanh(|q_{b}B|x_{2})}{|q_{b}B|(\tanh(|q_{b}B|x_{1})+\tanh(|q_{b}B|x_{2}))}s_{\perp}^{2}\right].\nonumber \\
    \label{Ilambda}
\end{eqnarray}
Next, $J(x_{1},x_{2})$ can be found using the change of variables
\begin{equation}
    q_{E\parallel}=k_{E\parallel}+\frac{x_{2}}{x_{1}+x_{2}}s_{E\parallel}. 
\end{equation}
Carrying out the integral and using $d=4-2\varepsilon$ we obtain
\begin{eqnarray}
    J(x_{1},x_{2})&=&\mu^{2\varepsilon}\left(\frac{1}{4\pi(x_1+x_2)}\right)^{1-\varepsilon}
    \nonumber\\
    &\times&e^{-\frac{x_1x_2}{x_1+x_2}s_{E\parallel}^{2}-(x_1+x_2)m_{\pi}^{2}}.
    \label{Jlambda}
\end{eqnarray}
Using the identities 
\begin{equation}
    \frac{e^{-x_{i}|q_{b}B|}}{1-r_{i}}=\frac{1}{2\cosh{(|q_{b}B|x_{i})}}, 
\end{equation}
\begin{equation}
\begin{split}
    \frac{1}{\sinh{(|q_{b}B|(x_{1}+x_{2}))}}&=\frac{1}{\tanh(|q_{b}B|x_{1})+\tanh(|q_{b}B|x_{2})}\\
    &\times \frac{1}{\cosh{(|q_{b}B|x_{1})}\cosh{(|q_{b}B|x_{2})}},   
\end{split}
\end{equation}    
together with Eqs.~\eqref{Ilambda} and \eqref{Jlambda}, we get
\begin{equation}
\begin{split}
    \Gamma^{B}_{\lambda}&=-\frac{\lambda}{12\pi^{2}}\int d^{2}x \frac{(4\pi \mu^{2})^{\varepsilon}}{(x_{1}+x_{2})^{1-\varepsilon}} \frac{|q_{b}B|}{\sinh{(|q_{b}B|(x_{1}+x_{2}))}} \\
    &\times \exp\left[-\frac{\tanh(|q_{b}B|x_{1})\tanh(|q_{b}B|x_{2})}{|q_{b}B|(\tanh(|q_{b}B|x_{1})+\tanh(|q_{b}B|x_{2}))}s_{\perp}^{2}\right] \\
    &\times\exp\left[-\frac{x_1x_2}{x_1+x_2}s_{E\parallel}^{2}-(x_1+x_2)m_{\pi}^{2}\right].
    \label{lambdaBdivergent}    
\end{split}
\end{equation}
We perform the change of variables
\begin{equation}
    x_{1}=s(1-y), \quad x_{2}=sy,\quad  dx_{1}dx_{2}=sdsdy.
    \label{change2}
\end{equation}
These variables have the domains $0<y<1$ and $s>0$. Substituting these new variables we obtain 
\begin{eqnarray}
    \Gamma^{B}_{\lambda}&=&-\frac{\lambda}{12\pi^{2}}\int_{0}^{\infty} ds \int_{0}^{1}dy\; (4\pi\mu^{2} s)^{\varepsilon} \frac{|q_{b}B|}{\sinh{(|q_{b}B|s)}} \nonumber\\
    &\times& \exp\left[-\frac{\tanh(|q_{b}B|s(1-y))\tanh(|q_{b}B|sy)}{|q_{b}B|(\tanh(|q_{b}B|s(1-y))+\tanh(|q_{b}B|sy))}s_{\perp}^{2}\right] \nonumber\\
    &\times&\exp\left[-sy(1-y) s_{E\parallel}^{2}-sm_{\pi}^{2}\right].
    \label{generallambda}   
\end{eqnarray}
Equation~\eqref{generallambda} is the general expression for the magnetic correction to the boson self-coupling. Notice that this expression contains a divergence that should be regularized. Considering the static limit in Eq.~\eqref{generallambda}, which implies that  $s_{E\parallel}^{2}\rightarrow 0$ and $s_{\perp}^{2}=0$, then the general magnetic correction reduces to 
\begin{equation}
    \Gamma^{B}_{\lambda}=-\frac{\lambda}{12\pi^{2}}\int_{0}^{\infty} ds \; (4\pi \mu^{2} s)^{\varepsilon} \frac{|q_{b}B|}{\sinh{(|q_{b}B|s)}} e^{-sm_{\pi}^{2}}.
\end{equation}
Notice that in this limit both integrals can be solved analytically
\begin{eqnarray}
   \int_{0}^{\infty}ds \frac{s^{\varepsilon}e^{-sm_{\pi}^{2}}}{\sinh{(|q_{b}B| s)}}&=&\frac{1}{|q_{b}B|}\left(\frac{1}{2|q_{b}B|}\right)^{\varepsilon}\Gamma(\varepsilon+1) \nonumber\\
   &\times&\zeta\left(\varepsilon+1,\frac{|q_{b}B| +m_{\pi}^{2}}{2|q_{b}B|} \right), 
\end{eqnarray}
where $\zeta$ is the Hurwitz zeta function. 
Considering an expansion for $\varepsilon\rightarrow 0$ we have
\begin{equation}
    \begin{split}
    &\left(\frac{4\pi \mu^{2}}{2|q_{b}B|}\right)^{\varepsilon}\approx 1+\varepsilon\ln{\left(\frac{4\pi \mu^{2}}{2|q_{b}B|} \right)},\\
    &\Gamma(\varepsilon+1)\approx 1-\varepsilon\gamma_{E},\\
    &\zeta\left(\varepsilon+1,\frac{|q_{b}B| +m_{\pi}^{2}}{2|q_{b}B|} \right)\approx \frac{1}{\varepsilon}-\psi^{0}\left(\frac{|q_{b}B|+m_{\pi}^{2}}{2|q_{b}B|}\right),          
    \end{split}
\end{equation}
where $\psi^{0}$ is the digamma function. Therefore, we finally obtain
\begin{eqnarray}
    \Gamma_{\lambda}^{B}&=&-\frac{\lambda}{12\pi^{2}} \bigg[\frac{1}{\varepsilon}-\gamma_{E}+\ln{(4\pi)}-\psi^{0}\left(\frac{|q_{b}B|+m_{\pi}^{2}}{2|q_{b}B|}\right)\nonumber\\
    &+&\ln{\left(\frac{\mu^{2}}{2|q_{b}B|} \right)}\bigg],
    \label{finalresultlambdaapp}
\end{eqnarray}
where $|q_{b}B|=|eB|$.
\section{Magnetic corrections to the boson-fermion coupling in the strong field limit }\label{computationmagneticg}
We start writing the contribution from the diagram in Fig.~\ref{fig:Feynmandiagramsg}(a) which can be obtained from the expression 
\begin{equation}
\begin{split}
I_{1,g}^{B}&=\int d^{4}x\;d^{4}y\;d^{4}x \int \frac{d^{4}s}{(2\pi)^{4}}\frac{d^{4}t}{(2\pi)^{4}}\frac{d^{4}k}{(2\pi)^{4}}\;e^{i\Phi_{1,l}} e^{-ip\cdot y}\\
&\times\left(\sqrt{2}g\gamma^{5} \right)e^{-is\cdot(x-y)}iS_{d}(s)\left(-g\gamma^{5}\right) e^{iq\cdot x} e^{-it\cdot(z-x)}\\
&\times iS_{d}(t)\left(\sqrt{2}g\gamma^{5} \right) e^{-ik\cdot(y-z)}iD_{\pi^{-}}(k)e^{ir\cdot z}+{\mbox{CC}},
\label{AppendixIg1}   
\end{split}
\end{equation}
where the Schwinger phase contribution is finite and is given by 
\begin{equation}
    \Phi_{1,l}=\frac{1}{2}|eB|\varepsilon_{ij}\left(x_{i}y_{j}+y_{i}z_{j}+z_{i}x_{j} \right),\quad i,j=1,2.
    \label{Schwingerphasetriangle1}
\end{equation}
The integration over configuration space can be performed using the factorization between parallel and perpendicular components. Recall that for four-vectors $a_{\mu}$ and $b_{\mu}$ then
\begin{equation}
    a_{\mu}b^{\mu}=a_{0}b_{0}-a_{1}b_{1}-a_{2}b_{2}-a_{3}b_{3}=a_{\parallel}\cdot b_{\parallel}-a_{\perp}\cdot b_{\perp}.
    \label{separationparper}
\end{equation}
Thus, integrating over configuration space and taking into account Eq.~\eqref{separationparper} to include the Schwinger phase contribution we obtain 
\begin{eqnarray}
I_{1,g}^{B}&=&\delta^{(2)}(p-q-r)_{\perp} \int  \frac{d^{4}s}{(2\pi)^{4}}\frac{d^{4}t}{(2\pi)^{4}}\frac{d^{4}k}{(2\pi)^{4}}\;\frac{4}{|eB|^2} (2\pi)^{10}\nonumber\\
& \times &\delta^{(2)}(s-q-t)_{\parallel}\;\delta^{(2)}(p-s+k)_{\parallel} \;\delta^{(2)}(t-k-r)_{\parallel}\nonumber\\  
&\times& \left(\sqrt{2}g\gamma^{5} \right)iS_{d}(s)\left(-g\gamma^{5}\right)iS_{d}(t) \left(\sqrt{2}g\gamma^{5} \right)iD_{\pi^{-}}(k)\nonumber\\
&\times& e^{i\frac{2}{|eB|}\varepsilon_{ij}(s-q-t)_{i}(s-p-k)_{j}}+{\mbox{CC}}.
\end{eqnarray}
We first integrate over $d^2s_\parallel$ and $d^2t_{\parallel}$ using the Dirac delta distributions to get
\begin{equation}
    \begin{split}
    I_{1,g}^{B}&=(2\pi)^{4} \delta^{(4)}(p-q-r) \int \frac{d^{2}s_{\perp} d^{2}t_{\perp}}{\pi^{2}|eB|^2}  \frac{d^{4}k}{(2\pi)^{4}}\left(\sqrt{2}g\gamma^{5} \right) \\ 
    &\times iS_{d}(k_{\parallel}+p_{\parallel},s_{\perp})\left(-g\gamma^{5}\right)iS_{d}(k_{\parallel}+r_{\parallel},t_{\perp})\left(\sqrt{2}g\gamma^{5} \right)\\
    &\times iD_{\pi^{-}}(k_{\parallel},k_{\perp})e^{i\frac{2}{|eB|}\varepsilon_{ij}(s-q-t)_{i}(s-p-k)_{j}}+{\mbox{CC}}.
    \label{resultI1g}
    \end{split} 
\end{equation}
Notice that with this procedure we can identify the Dirac delta distribution for energy-momentum conservation in Eq.~\eqref{resultI1g} such that
\begin{equation}
    I_{1,g}^{B}=(2\pi)^{4}\delta^{(4)}(p-r-q)g\gamma^{5}\Gamma_{1,g}^{B}.
\end{equation}
The contribution to the magnetic correction to the boson-fermion coupling, $g\gamma^{5}\Gamma_{1,g}^{B}$, is thus given by
\begin{eqnarray}
   g\gamma^{5}\Gamma_{1,g}^{B}&=&\int \frac{d^{2}s_{\perp} d^{2}t_{\perp}}{\pi^{2}|eB|^2} \frac{d^{4}k}{(2\pi)^{4}}\left(\sqrt{2}g\gamma^{5} \right)iS_{d}(k_{\parallel}+p_{\parallel},s_{\perp}) \nonumber\\
   &\times&\left(-g\gamma^{5}\right) iS_{d}(k_{\parallel}+r_{\parallel},t_{\perp})\left(\sqrt{2}g\gamma^{5} \right)iD_{\pi^{-}}(k_{\parallel},k_{\perp})\nonumber\\ &\times&  e^{i\frac{2}{|eB|}\varepsilon_{ij}(s-q-t)_{i}(s-p-k)_{j}}+ {\mbox{CC}}. 
\label{AppendixGammag1}
\end{eqnarray}
Equation~\eqref{AppendixGammag1} is general enough and could be computed using either the complete propagators or approximations to them. In this work we consider the propagators in the strong field limit. Substituting Eqs.~\eqref{LLLbos} and \eqref{fermionpropagatorLLL} and adding the charge conjugate contribution we have
\begin{eqnarray}
&&\Gamma_{1,g}^{LLL}= \frac{16ig^{2}}{\pi^{2}|eB|^2}\int d^{2}s_{\perp} d^{2}t_{\perp} \frac{d^{4}k}{(2\pi)^{4}} e^{-\frac{s_{\perp}^{2}}{|q_{d}B|}-\frac{t_{\perp}^{2}}{|q_{d}B|}-\frac{k_{\perp}^{2}}{|q_{\pi^{-}}B|}} \nonumber \\
&&\times\frac{\mathcal{N}_{1}}{A_{1}B_{1}C_{1}} \; e^{i\frac{2}{|eB|}\varepsilon_{ij}(s-q-t)_{i}(s-p-k)_{j}},
\end{eqnarray}
where we have defined for convenience the quantities
\begin{eqnarray}
&&\mathcal{N}_{1}= (\slashed{k}_{\parallel}+\slashed{p}_{\parallel}+m_{d})(m_{d}-\slashed{k}_{\parallel}-\slashed{r}_{\parallel}),  \nonumber\\
&&A_{1}=(k_{\parallel}+p_{\parallel})^{2}-m_{d}^{2}+i\epsilon, \nonumber \\
&&B_{1}=(k_{\parallel}+r_{\parallel})^{2}-m_{d}^{2}+i\epsilon, \nonumber\\
&&C_{1}=k_{\parallel}^{2}-m_{\pi}^{2}-|eB|+i\epsilon.   
\end{eqnarray}
We also resort to work in the static limit, setting the perpendicular coordinates of external momenta to zero. On doing so, we can integrate over the perpendicular coordinates relative to the magnetic field. The result is given by 
\begin{equation}
\Gamma_{1,g}^{LLL}= \frac{ig^{2}|eB|}{4\pi} \int \frac{d^{2}k_{\parallel}}{(2\pi)^{2}}\frac{\mathcal{N}_{1}}{A_{1}B_{1}C_{1}},
\end{equation}
Introducing the Feynman parametrization
\begin{equation}
    \frac{1}{A_{1}B_{1}C_{1}}=\int_{0}^{1}dx\int_{0}^{1-x} \frac{2dy}{(A_{1}x+B_{1}y+C_{1}(1-x-y))^{3}}.
    \label{twoFeynmanparameters}
\end{equation}
The denominator of Eq.~(\ref{twoFeynmanparameters}) can be expressed as
\begin{equation}
    A_{1}x+B_{1}y+C_{1}(1-x-y)=(k_{\parallel}+xp_{\parallel}+yr_{\parallel})^{2}-\Delta+i\epsilon,
\end{equation}
where 
\begin{eqnarray}
    \Delta &=&(xp_{\parallel}+yr_{\parallel})^{2}-xp_{\parallel}^{2}+xm_{d}^{2}-yr_{\parallel}^{2}+ym_{d}^{2}\nonumber \\
    &+&(1-x-y)(m_{\pi}^{2}+|eB|).
\end{eqnarray}
On the other hand, it is useful to consider the change of variables as  $k_{\parallel}=l_{\parallel}-xp_{\parallel}-yr_{\parallel}$, $dk_{\parallel}=dl_{\parallel}$. Then the numerator, $\mathcal{N}$, can be written as
\begin{eqnarray}  \mathcal{N}_{1}&=&-l_{\parallel}^{2}-2xyp_{\parallel}\cdot r_{\parallel}+m_{d}\slashed{p}_{\parallel}-m_{d}\slashed{r}_{\parallel}-x(x-1)p_{\parallel}^{2}\nonumber\\
&-&y(y-1)r_{\parallel}^{2}-(1-x-y)\slashed{p}_{\parallel}\slashed{r}_{\parallel}+m_{d}^{2},
\end{eqnarray}
where we have already discarded linear terms of $l_{\parallel}$. 
At this point we can use the Dirac equation for outgoing states assuming that they are not affected by the external magnetic field. This means that the spinors satisfy the Dirac equation in vacuum
\begin{equation}
    \bar{u}(p_{\parallel})\slashed{p}_{\parallel}=\bar{u}(p_{\parallel})m_{u}, \quad \slashed{r}_{\parallel}u(r_{\parallel})=m_{u}u(r_{\parallel}).
    \label{DiracEquation}
\end{equation}
Here, it is worth to note that in this computation we assume that the values of the quark masses remain fixed to just their vacuum values, $m_{d}=m_{u}=m_{f}$. Then, taking the static limit, $p_{3}=r_{3}=0$ and $p_{0}=r_{0}=m_{f}$, we get
\begin{equation}
     \bar{u}(p_{\parallel})\mathcal{N}_{1}u(r_{\parallel})\!\!=\!\!\bar{u}(p_{\parallel})(-l_{\parallel}^{2}+2m_{f}^{2}(x+y)-m_{f}^{2}(x+y)^{2})u(r_{\parallel}).
     \label{numerator1withDirac}
\end{equation}
Thus, once we consider $\bar{u}(p_{\parallel})\Gamma_{1,g}^{LLL} u(r_{\parallel})$ and use Eq.~\eqref{numerator1withDirac} we get
\begin{eqnarray}
    \Gamma_{1,g}^{LLL}&=& \frac{ig^{2}|eB|}{2\pi}\int_{0}^{1}dx\int_{0}^{1-x}dy\int \frac{d^{2}l_{\parallel}}{(2\pi)^{2}}\bigg[
    \frac{-l_{\parallel}^{2}}{(l_{\parallel}^{2}-\Delta+i\epsilon)^{3}}\nonumber\\
    &+&\frac{2m_{f}^{2}(x+y)-m_{f}^{2}(x+y)^{2}}{(l_{\parallel}^{2}-\Delta+i\epsilon)^{3}}\bigg],
\end{eqnarray}
where with the above assumptions, $\Delta$ is simplified to become
\begin{equation}
    \Delta=m_{f}^{2}(x+y)^{2}+(1-(x+y))(m_{\pi}^{2}+|eB|). 
\end{equation}
In order to integrate over $d^{2}l_{\parallel}$ we consider the following equations
\begin{equation}
    \mu^{4-d}\int \frac{d^{d-2}l_{\parallel}}{(2\pi)^{d-2}}\frac{1}{(l_{\parallel}^{2}-\Delta)^{3}}=-\frac{i}{4\pi}\frac{1}{2}\frac{1}{\Delta^{2}}+\mathcal{O}(\varepsilon),
    \label{dimensionalregularizationparallel3}
\end{equation}
\begin{equation}
    \mu^{4-d}\int \frac{d^{d-2}l_{\parallel}}{(2\pi)^{d-2}}\frac{l_{\parallel}^{2}}{(l_{\parallel}^{2}-\Delta)^{3}}=\frac{i}{4\pi}\frac{1}{2}\frac{1}{\Delta}+\mathcal{O}(\varepsilon).
    \label{dimensionalregularizationparallel4}
\end{equation}
According to Eqs. \eqref{dimensionalregularizationparallel3} and \eqref{dimensionalregularizationparallel4} we get
\begin{eqnarray}
    \Gamma_{1,g}^{LLL}&=&\frac{g^{2}|eB|}{16\pi^{2}m_{f}^{2}}\int_{0}^{1}dx\int_{0}^{1-x}dy \bigg[\frac{1}{(x+y)^{2}+\alpha (1-(x+y))}\nonumber\\
    &+&\frac{2(x+y)-(x+y)^{2}}{ ((x+y)^{2}+\alpha(1-(x+y)))^{2}} \bigg],
\end{eqnarray}
where $\alpha=(m_{\pi}^{2}+|eB|)/m_{f}^{2}$. With the purpose of finding the integral over Feynman parameters, consider the following linear transformation
\begin{equation}
    u=x+y, \quad v=1-x.
    \label{changecoordinates}
\end{equation}
The Jacobian  satisfies $det(J)=1$
and the region of integration becomes $u\in[0,1]$ and $v\in[1-u,1]$. Thus,
\begin{eqnarray}
    \Gamma_{1,g}^{LLL}&=&\frac{g^{2}|eB|}{16\pi^{2}m_{f}^{2}}\int_{0}^{1}du\int_{1-u}^{1}dv \frac{1}{u^{2}+\alpha (1-u)} \nonumber\\
    &\times& \left[1+\frac{(2-u)u}{u^{2}+\alpha (1-u)} \right].   
\end{eqnarray}
Performing the integration over $dv$ we get the final expression for this contribution
\begin{eqnarray}
    \Gamma_{1,g}^{LLL}&=&\frac{g^{2}|eB|}{16\pi^{2}m_{f}^{2}}\int_{0}^{1}du\frac{u}{ u^{2}+\alpha(1-u)}\nonumber\\
    &\times&\left[1+\frac{(2-u)u}{u^{2}+\alpha(1-u)} \right].
    \label{AppendixresultGammag1}
\end{eqnarray}
We now proceed with $I_{2,g}$ which is given by
\begin{equation}
\begin{split}
I_{2,g}^{B}&=\int d^{4}x\;d^{4}y\;d^{4}z \int \frac{d^{4}s}{(2\pi)^{4}}\frac{d^{4}t}{(2\pi)^{4}}\frac{d^{4}k}{(2\pi)^{4}}e^{i\Phi_{2,l}} e^{-ip\cdot y}\\
&\times\left(g\gamma^{5} \right)e^{-is\cdot(x-y)}iS_{u}(s)\left(g\gamma^{5}\right)e^{iq\cdot x} e^{-it\cdot(z-x)}\\ 
& \times   iS_{u}(t)\left(g\gamma^{5} \right) e^{-ik\cdot(y-z)}iD_{\pi^{0}}(k)e^{ir\cdot z}+{\mbox{CC}}.    
\end{split}
\end{equation}
Performing the integration over configuration space we have
\begin{eqnarray}
I_{2,g}^{B}&=& \int \frac{d^{4}s}{(2\pi)^{4}}\frac{d^{4}t}{(2\pi)^{4}}\frac{d^{4}k}{(2\pi)^{4}}(2\pi)^{12}\delta^{(4)}(s-t-q) \nonumber\\
&\times& \delta^{(4)}(p-s+k)\delta^{(4)}(t-k-r) \left(g\gamma^{5} \right)iS_{u}(s) \nonumber \\ 
&\times& \left(g\gamma^{5}\right)iS_{u}(t)\left(g\gamma^{5} \right) iD_{\pi^{0}}(k)+{\mbox{CC}},
\end{eqnarray}
Integrating over $d^{4}s$ and $d^{4}t$ we obtain
\begin{eqnarray}
I_{2,g}^{B}&=&(2\pi)^{4}\delta^{(4)}(p-r-q) \int\frac{d^{4}k}{(2\pi)^{4}} \left(g\gamma^{5} \right)iS_{u}(k+p) \nonumber\\
&\times&\left(g\gamma^{5}\right)iS_{u}(k+r)\left(g\gamma^{5} \right) iD_{\pi^{0}}(k)+{\mbox{CC}}.
\label{resultIg2}
\end{eqnarray}
At this point we can identify the contribution to the magnetic correction from this diagram, $g\gamma^{5}\Gamma_{2,g}^{B}$, which can be expressed as
\begin{equation}
    I_{2,g}^{B}=(2\pi)^{4}\delta^{(4)}(p-r-q)g\gamma^{5}\Gamma_{2,g}^{B}, 
\end{equation}
where 
\begin{eqnarray}
    g\gamma^{5}\Gamma_{2,g}^{B}&=&\int\frac{d^{4}k}{(2\pi)^{4}} \left(g\gamma^{5} \right)iS_{u}(k+p)\left(g\gamma^{5}\right)iS_{u}(k+r) \nonumber\\
    &\times&\left(g\gamma^{5} \right) iD_{\pi^{0}}(k)+{\mbox{CC}},
    \label{AppendixGammag2}
\end{eqnarray}
using Eqs.~\eqref{LLLbos} and~\eqref{fermionpropagatorLLL} to account for the strong field limit, we have
\begin{equation}
    \Gamma_{2,g}^{LLL}=-4ig^{2}\int\frac{d^{4}k}{(2\pi)^{4}}e^{-\frac{(k+p)_{\perp}^{2}}{|q_{u}B|}-\frac{(k+r)_{\perp}^{2}}{|q_{u}B|}} \frac{\mathcal{N}_{2}}{A_{2}B_{2}C_{2}},
\end{equation}
where we define 
\begin{eqnarray}
    &&\mathcal{N}_{2}= (\slashed{k}_{\parallel}+\slashed{p}_{\parallel}+m_{u})(m_{u}-\slashed{k}_{\parallel}+\slashed{r}_{\parallel}),\nonumber\\
    &&A_{2}=(k_{\parallel}+p_{\parallel})^{2}-m_{u}^{2}+i\epsilon,\nonumber\\
    &&B_{2}=(k_{\parallel}+r_{\parallel})^{2}-m_{u}^{2}+i\epsilon,\nonumber\\
    &&C_{2}=k^{2}-m_{\pi}^{2}+i\epsilon.   
\end{eqnarray}
We now introduce a Feynman parametrization in the same fashion of Eq.~\eqref{twoFeynmanparameters}. The denominator can be written as 
\begin{equation}
    A_{2}x+B_{2}y+C_{2}(1-x-y)=(k_{\parallel}+xp_{\parallel}+yr_{\parallel})^{2}-\Delta_{
    \perp}+i\epsilon,
\end{equation}
where
\begin{eqnarray}
    \Delta_{\perp} &=&(xp_{\parallel}+yr_{\parallel})^{2}-xp_{\parallel}^{2}+(x+y)m_{u}^{2}-yr_{\parallel}^{2}\nonumber\\
    &+&(1-x-y)(m_{\pi}^{2}+k_{\perp}^{2}).
\end{eqnarray}
Let us consider the change of variable $k_{\parallel}=l_{\parallel}-xp_{\parallel}-yr_{\parallel}$, $dk_{\parallel}=dl_{\parallel}$, then in terms of these variables, the numerator $\mathcal{N}_{2}$ can be written as
\begin{eqnarray}
    \mathcal{N}_{2}&=&-l_{\parallel}^{2}-2xyp_{\parallel}\cdot r_{\parallel}+m_{u}\slashed{p}_{\parallel}-m_{u}\slashed{r}_{\parallel}-x(x-1)p_{\parallel}^{2} \nonumber\\
    &-&y(y-1)r_{\parallel}^{2}-(1-x-y)\slashed{p}_{\parallel}\slashed{r}_{\parallel}+m_{u}^{2},
\end{eqnarray}
where already discarded linear terms in $l_{\parallel}$. We now use the Dirac equation for outgoing states once we set $p_{i}=r_{i}=0,$ $i=1,2$ and assume that these states are not affected by the external magnetic field, according to Eq.~\eqref{DiracEquation}. Finally, setting $p_{3}=r_{3}=0$ and $p_{0}=r_{0}=m_{u}$ we get
\begin{equation}
     \bar{u}(p_{\parallel})\mathcal{N}_{2}u(r_{\parallel})\!\!=\!\!\bar{u}(p_{\parallel})(-l_{\parallel}^{2}+2m_{u}^{2}(x+y)-m_{u}^{2}(x+y)^{2})u(r_{\parallel}).
\end{equation}
Thus, once we have considered $\bar{u}(p_{\parallel})\Gamma_{2,g}^{LLL}u(r_{\parallel})$ we have
\begin{equation}
\begin{split}
    \Gamma_{2,g}^{LLL}&=-8ig^{2}\int_{0}^{1}dx\int_{0}^{1-x}dy\int\frac{d^{2}k_{\perp}d^{2}l_{\parallel}}{(2\pi)^{4}}
    e^{-\frac{2k_{\perp}^{2}}{|q_{u}B|}} \\
    &\times \left[\frac{-l_{\parallel}^{2}}{(l_{\parallel}^{2}-\Delta_{\perp}+i\epsilon)^{3}}+\frac{2m_{u}^{2}(x+y)-m_{u}^{2}(x+y)^{2}}{(l_{\parallel}^{2}-\Delta_{\perp}+i\epsilon)^{3}}\right],
\end{split}
\end{equation}
where $\Delta_{\perp}$ is simplified according to the previous assumptions to become
\begin{equation}
    \Delta_{\perp}=m_{u}^{2}(x+y)^{2}+(1-(x+y))(k_{\perp}^{2}+m_{\pi}^{2}). 
\end{equation}
The integral over $d^{2}l_{\parallel}$ is found to be
\begin{eqnarray}
    \Gamma_{2,g}^{LLL}&=&-\frac{g^{2}}{\pi}\int_{0}^{1}dx\int_{0}^{1-x}dy\int \frac{d^{2}k_{\perp}}{(2\pi)^{2}}
    e^{-\frac{2k_{\perp}^{2}}{|q_{u}B|}}\nonumber\\
    &\times&\left[\frac{1}{ \Delta_{\perp}}+\frac{2m_{u}^{2}(x+y)-m_{u}^{2}(x+y)^{2}}{ \Delta_{\perp}^{2}} \right].
\end{eqnarray}
Using the change of variables given in Eq.~\eqref{changecoordinates}  the integral over $dv$ can be performed to get
\begin{eqnarray}
   \Gamma_{2,g}^{LLL}&=&-\frac{g^{2}}{\pi m_{u}^{2}}\int_{0}^{1}du\int \frac{d^{2}k_{\perp}}{(2\pi)^{2}}
    e^{-\frac{2k_{\perp}^{2}}{|q_{u}B|}}\nonumber \\
    &\times& \frac{u}{ u^{2}+\beta(1-u)}\left[1+\frac{(2-u)u}{ u^{2}+\beta(1-u)} \right],
\end{eqnarray}
where $\beta = (k_{\perp}^{2}+m_{\pi}^{2})/m_{u}^{2}$. We write the integration using polar coordinates
\begin{equation}
    d^{2}k_{\perp}=dk_{1}dk_{2}=k_{\perp}dk_{\perp} d\theta,
    \label{polarcoordinates}
\end{equation}
where $k_{\perp}=\sqrt{k_{1}^{2}+k_{2}^{2}}$ and $\theta \in [0,2\pi]$. Performing the integral over $d\theta$ and substituting $|q_{u}B|=2|eB|/3$ and $m_{u}=m_{f}$ we have
\begin{eqnarray}
   \Gamma_{2,g}^{LLL}&=&-\frac{g^{2}}{2\pi^{2} m_{f}^{2}}\int_{0}^{1}du \int_{0}^{\infty} dk_{\perp}\; k_{\perp} e^{-\frac{3k_{\perp}^{2}}{|eB|}}\nonumber\\
   &\times&\frac{u}{ u^{2}+\beta(1-u)}\left[1+\frac{(2-u)u}{ u^{2}+\beta(1-u)} \right].
   \label{AppendixresultGammag2}
\end{eqnarray}
Finally, $I_{3,g}^{B}$ can be written as
\begin{eqnarray}
I_{3,g}^{B}&=&\int d^{4}x\;d^{4}y\;d^{4}z \int \frac{d^{4}s}{(2\pi)^{4}}\frac{d^{4}t}{(2\pi)^{4}}\frac{d^{4}k}{(2\pi)^{4}}e^{i\Phi_{3,l}} e^{-ip\cdot y} \nonumber \\
&\times&\left(-ig \right)e^{-is\cdot(x-y)}iS_{u}(s)\left(g\gamma^{5}\right) e^{iq\cdot x} e^{-it\cdot(z-x)} \nonumber\\
&\times& iS_{u}(t)\left(-ig \right) e^{-ik\cdot(y-z)}iD_{\sigma}(k)e^{ir\cdot z}+{\mbox{CC}}.
\label{AppendixIg3}
\end{eqnarray}
After integration over configuration space we get
\begin{eqnarray}
I_{3,g}^{B}&=& \int \frac{d^{4}s}{(2\pi)^{4}}\frac{d^{4}t}{(2\pi)^{4}}\frac{d^{4}k}{(2\pi)^{4}} (2\pi)^{12}\delta^{(4)}(s-t-q)\nonumber \\
&\times&\delta^{(4)}(p-s+k)\delta^{(4)}(t-k-r)\left(-ig\right)iS_{u}(s) \nonumber\\ 
&\times& \left(g\gamma^{5}\right)iS_{u}(t)\left(-ig \right) iD_{\sigma}(k)+\mbox{CC},
\end{eqnarray}
integrating over $d^{4}s$ and $d^{4}t$ we have
\begin{eqnarray}
I_{3,g}^{B}&=&(2\pi)^{4}\delta^{(4)}(p-q-r)\int \frac{d^{4}k}{(2\pi)^{4}} \left(-ig\right)iS_{u}(k+p) \nonumber\\
&\times&\left(g\gamma^{5}\right)iS_{u}(k+r)\left(-ig \right) iD_{\sigma}(k)+\mbox{CC},
\label{resultIg3}
\end{eqnarray}
from where we can identify the contribution to the magnetic correction according to the expression
\begin{equation}
    I_{3,g}^{B}=(2\pi)^{4}\delta^{(4)}(p-r-q)g\gamma^{5}\Gamma_{3,g}, 
\end{equation}
where
\begin{eqnarray}
g\gamma^{5}\Gamma_{3,g}^{B}&=& \int \frac{d^{4}k}{(2\pi)^{4}} \left(-ig \right)iS_{u}(k+p)\left(g\gamma^{5}\right)iS_{u}(k+r)\nonumber\\
&\times&\left(-ig \right) iD_{\sigma}(k)+{\mbox{CC}}.
\end{eqnarray}
We now use the propagators for the charged particles in the LLL. After simplifying and adding the contribution from the charge conjugate diagram we get
\begin{equation}
    \Gamma_{3,g}^{LLL}\!=\!4ig^{2} \int \frac{d^{4}k}{(2\pi)^{4}} e^{-\frac{(k+p)_{\perp}^{2}}{|q_{u}B|}-\frac{(k+r)_{\perp}^{2}}{|q_{u}B|}}\frac{\mathcal{N}_{3}}{A_{3}B_{3}C_{3}},
\end{equation}
where we define 
\begin{eqnarray}
    &&\mathcal{N}_{3}=(m_{u}-\slashed{k}_{\parallel}-\slashed{p}_{\parallel})(\slashed{k}_{\parallel}+\slashed{r}_{\parallel}+m_{u}),\nonumber\\
    &&A_{3}=(k_{\parallel}+p_{\parallel})^{2}-m_{u}^{2}+i\epsilon, \nonumber\\
    &&B_{3}=(k_{\parallel}+r_{\parallel})^{2}-m_{u}^{2}+i\epsilon, \nonumber\\
    &&C_{3}=k^{2}-m_{\sigma}^{2}+i\epsilon.   
\end{eqnarray}
The denominator can be written as
\begin{equation}
    A_{3}x+B_{3}y+C_{3}(1-x-y)=(k_{\parallel}+xp_{\parallel}+yr_{\parallel})^{2}-\Delta_{
    \perp}+i\epsilon,
\end{equation}
where
\begin{eqnarray}
\Delta_{\perp} &=&(xp_{\parallel}+yr_{\parallel})^{2}-xp_{\parallel}^{2}+(x+y)m_{u}^{2}-yr_{\parallel}^{2}\nonumber\\
&+&(1-x-y)(m_{\sigma}^{2}+k_{\perp}^{2}).
\end{eqnarray}
Using the change of variable $k_{\parallel}=l_{\parallel}-xp_{\parallel}-yr_{\parallel}$, $dk_{\parallel}=dl_{\parallel}$,  the numerator, $\mathcal{N}_{3}$, can be written as
\begin{eqnarray}
    \mathcal{N}_{3}&=&-l_{\parallel}^{2}-2xyp_{\parallel}\cdot r_{\parallel}-m_{u}\slashed{p}_{\parallel}+m_{u}\slashed{r}_{\parallel}-x(x-1)p_{\parallel}^{2}\nonumber\\
    &-&y(y-1)r_{\parallel}^{2}-(1-x-y)\slashed{p}_{\parallel}\slashed{r}_{\parallel}+m_{u}^{2},
\end{eqnarray}
where we have neglected linear terms of $l_{\parallel}$. We proceed as for the previous cases. We use Eq.~\eqref{DiracEquation} and work in the static limit, $\vec{p}=\vec{r}=\vec{0}$ and $p_{0}=r_{0}=m_{u}$, to obtain
\begin{equation}
     \bar{u}(p_{\parallel})\mathcal{N}_{3}u(r_{\parallel})\!\!=\!\!\bar{u}(p_{\parallel})(-l_{\parallel}^{2}+2m_{u}^{2}(x+y)-m_{u}^{2}(x+y)^{2})u(r_{\parallel}).
\end{equation}
Thus, the integral can be written as
\begin{equation}
\begin{split}    
     \Gamma_{3,g}^{LLL}&=8ig^{2} \int \frac{d^{2}k_{\perp}}{(2\pi)^{2}}\int_{0}^{1}dx\int_{0}^{1-x}dy \int \frac{d^{2}l_{\parallel}}{(2\pi)^{2}}\; e^{-\frac{2k_{\perp}^{2}}{|q_{u}B|}} \\
     &\times\left[\frac{-l_{\parallel}^{2}}{(l_{\parallel}^{2}-\Delta_{
    \perp}+i\epsilon)^{3}}+ 
     \frac{2m_{u}^{2}(x+y)-m_{u}^{2}(x+y)^{2}}{(l_{\parallel}^{2}-\Delta_{
    \perp}+i\epsilon)^{3}}\right],
\end{split}    
\end{equation}
where 
\begin{equation}
\Delta_{\perp}=m_{u}^{2}(x+y)^{2}+(1-(x+y))(k_{\perp}^{2}+m_{\sigma}^{2}).
\end{equation}
Now, we can perform the integration over $d^{2}l_{\parallel}$ to get
\begin{eqnarray}
   \Gamma_{3,g}^{LLL}&=&\frac{g^{2}}{\pi}\int \frac{d^{2}k_{\perp}}{(2\pi)^{2}}\int_{0}^{1}dx\int_{0}^{1-x}dy
    e^{-\frac{2k_{\perp}^{2}}{|q_{u}B|}} \nonumber\\
    &\times&\left[\frac{1}{ \Delta_{\perp}}+\frac{2m_{u}^{2}(x+y)-m_{u}^{2}(x+y)^{2}}{ \Delta_{\perp}^{2}} \right].
\end{eqnarray}
The last expression can be simplified if we consider the change of variables given by Eq.~\eqref{changecoordinates}. After integration over $dv$ we have
\begin{eqnarray}
   \Gamma_{3,g}^{LLL}&=&\frac{g^{2}}{\pi  m_{u}^{2}}\int_{0}^{1}du \int \frac{d^{2}k_{\perp}}{(2\pi)^{2}}\;
    e^{-\frac{2k_{\perp}^{2}}{|q_{u}B|}}  \nonumber\\
    &\times&\frac{u}{u^{2}+\gamma(1-u)}\left[1+\frac{(2-u)u}{u^{2}+\gamma(1-u)} \right],
\end{eqnarray}
where $\gamma=(k_{\perp}^{2}+m_{\sigma}^{2})/m_{u}^{2}$. We can now perform another integration after switching to polar coordinates according to Eq.~(\ref{polarcoordinates}). Performing the integration for $d\theta$ and substituting $|q_{u}B|=2|eB|/3$ and $m_{u}=m_{f}$ we have the final result
\begin{eqnarray}
   \Gamma_{3,g}^{LLL}&=&\frac{g^{2}}{2\pi^{2}m_{f}^{2}}\int_{0}^{1}du \int_{0}^{\infty} dk_{\perp}\;
    k_{\perp}e^{-\frac{3k_{\perp}^{2}}{|eB|}} \nonumber\\
    &\times&\frac{u}{u^{2}+\gamma(1-u)}\left[1+\frac{(2-u)u}{u^{2}+\gamma(1-u)} \right].
\end{eqnarray}
\end{appendix}



\begin{thebibliography}{89}

\bibitem{LQCD}
G. S. Bali, F. Bruckmann, G. Endr\"odi, Z. Fodor, S. D. Katz, S. Krieg, A. Sch\"afer and K. K. Szab\'o, J. High Energy Phys. {\bf 02} (2012) 044; G. S. Bali, F. Bruckmann, G. Endr\"odi, Z. Fodor, S. D. Katz and A. Sch\"afer, Phys. Rev. D {\bf 86}, 071502 (2012); G. Bali, F. Bruckmann, G. Endr\"odi, S. Katz and A. Sch\"afer, J. High Energy Phys. {\bf 08} (2014) 177.

\bibitem{Bruckmann}
F. Bruckmann, G. Endr\"odi and T. G. Kovacs, J. High Energy Phys. {\bf 04} (2013) 112.

\bibitem{Farias} R. L. S. Farias, K. P. Gomes, G. Krein and M. B. Pinto, Phys. Rev. C {\bf 90}, 025203 (2014).

\bibitem{Ferreira} M. Ferreira, P. Costa, O. Louren\c{c}o, T. Frederico and C. Provid\^encia, Phys. Rev. D {\bf 89}, 116011 (2014).

\bibitem{Ayala0}
A. Ayala, L. A. Hernández, A. J. Mizher, J. C. Rojas and C. Villavicencio, Phys. Rev. D {\bf 89}, 116017  (2014).

\bibitem{Ayala1} A. Ayala, M. Loewe and R. Zamora, Phys. Rev. D {\bf 91}, 016002 (2015).

\bibitem{Ayala2} A. Ayala, C. A. Dominguez, L. A.
Hern\'andez, M. Loewe and R. Zamora, Phys. Rev. D {\bf 92}, 096011 (2015).

\bibitem{Ayala3} A. Ayala, M. Loewe, A. J. Mizher and R. Zamora, Phys. Rev. D {\bf 90}, 036001 (2014).

\bibitem{Avancini} R. L. S. Farias, V. S. Timoteo, S. S. Avancini, M. B. Pinto and G. Krein, Eur. Phys. J. A {\bf 53}, 101 (2017)

\bibitem{Ayala4} A. Ayala, C. A. Dominguez, L. A. Hern\'andez, M. Loewe, A. Raya, J. C. Rojas and C. Villavicencio, Phys. Rev. D {\bf 94}, 054019 (2016).

\bibitem{vertex1}
A. Ayala, C. A. Dominguez, L. A. Hern\'andez, M. Loewe and R. Zamora, Phys. Lett. B {\bf 759}, 99-103 (2016).

\bibitem{vertex2} A. Ayala, J. J. Cobos-Mart\'inez, M. Loewe, M. E. Tejeda-Yeomans and R. Zamora, Phys. Rev. D {\bf 91}, 016007 (2015).

\bibitem{qcdcoupling} A. Ayala, C. A. Dominguez, S. Hern\'andez-Ort\'iz, L. A. Hern\'andez, M. Loewe, D. Manreza Paret and R. Zamora, Phys. Rev. D {\bf 98}, 031501 (2018).

\bibitem{Mueller} 	N. Mueller, J. A. Bonnet and C. S. Fischer, Phys. Rev. D {\bf 89}, 094023 (2014); N. Mueller and J. M. Pawlowski, Phys. Rev. D {\bf 91}, 116010  (2015).

\bibitem{imcreview} A. Bandyopadhyay and R.L.S. Farias, arXiv:2003.11054 [hep-ph].

\bibitem{bali01} G. S. Bali, B. B. Brandt, G. Endr{\H o}di and B. Gl{\"a}ssle, Phys. Rev. Lett. {\bf 121}, 072001 (2018).
%
\bibitem{iranianos} Sh. Fayazbakhsh and N. Sadooghi, Phys. Rev. D {\bf 88}, 065030 (2013).
%
\bibitem{simonov03} Yu. A. Simonov, Phys. At. Nucl. {\bf 79}, 455 (2016).
%
\bibitem{aguirre02} R.~M.~Aguirre,
Eur. Phys. J. A \textbf{55}, 28 (2019).
%
\bibitem{tetsuya} T. Yoshida and K. Suzuki,
Phys. Rev. D {\bf 94}, 074043 (2016).
%
\bibitem{dudal04} D. Dudal and T. G. Mertens,
Phys. Rev. D {\bf 91}, 086002 (2015).
%
\bibitem{kevin} K.~Marasinghe and K.~Tuchin,
Phys. Rev. C {\bf 84}, 044908 (2011).
%
\bibitem{gubler} P. Gubler, K. Hattori, S. H. Lee, M. Oka, S. Ozaki and K. Suzuki, Phys. Rev. D {\bf 93}, 054026 (2016).
%
\bibitem{noronha01} C. S. Machado, S. I. Finazzo, R. D. Matheus and J. Noronha, Phys. Rev. D {\bf 89}, 074027 (2014).
%
\bibitem{morita} S. Cho, K. Hattori, S. H. Lee, K. Morita and S. Ozaki,
Phys. Rev. Lett. {\bf 113}, 172301 (2014).
%
\bibitem{morita02}  S. Cho, K. Hattori, S. H. Lee, K. Morita and S. Ozaki,
Phys. Rev. D {\bf 91}, 045025 (2015).
%
\bibitem{sarkar03} S.~Ghosh, A.~Mukherjee, M.~Mandal, S.~Sarkar and P.~Roy,
Phys. Rev. D \textbf{94}, 094043 (2016).
%
\bibitem{band} A.~Bandyopadhyay and S.~Mallik,
Eur. Phys. J. C \textbf{77}, 771 (2017).
%
\bibitem{Ayalachi}
A. Ayala, L. A. Hern\'andez, A. J. Mizher, J. C. Rojas, C. Villavicencio, Phys. Rev. D {\bf 89}, 116017 (2014).
%
\bibitem{nosso1} S. S. Avancini, W. R. Tavares and M. B. Pinto, Phys. Rev. D {\bf 93}, 014010 (2016).
%
\bibitem{nosso03} S. S. Avancini, R. L. Farias, M. B. Pinto, W. R. Tavares and V. S. Timteo, Phys. Lett. B {\bf 767}, 247 (2017).
%
\bibitem{zhuang} Z. Wang and P. Zhuang, Phys. Rev. D {\bf 97}, 034026 (2018).
%
\bibitem{iran} Sh. Fayazbakhsh, S. Sadeghian and N. Sadooghi, Phys. Rev. D {\bf 86}, 085042 (2012). 
%
\bibitem{scoccola01} M. Coppola, D. G{\'o}mez Dumm and N. N. Scoccola, Phys. Lett. B {\bf 782}, 155 (2018).
%
\bibitem{huang01} H. Liu, X. Wang, L. Yu and M. Huang, Phys. Rev. D {\bf 97}, 076008 (2018).
%
\bibitem{scoccola02} D. G{\'o}mez Dumm, M. F. Izzo Villafa{\~n}e and N. N. Scoccola,
Phys. Rev. D {\bf 97}, 034025 (2018).
%
\bibitem{scoccola03} D.~G\'omez Dumm, M.~F.~Izzo Villafa\~ne and N.~N.~Scoccola,
Phys. Rev. D \textbf{101}, no.11, 116018 (2020).
%
\bibitem{luch} M.~A.~Andreichikov, B.~O.~Kerbikov, E.~V.~Luschevskaya, Y.~A.~Simonov and O.~E.~Solovjeva,
J. High Energy Phys. \textbf{05}, 007 (2017).
%
\bibitem{farias01} S. S. Avancini, R. L. S. Farias and W. R. Tavares, Phys. Rev. D {\bf 99}, 056009 (2019).
%
\bibitem{mao01} S.~Mao,
Phys. Rev. D \textbf{99}, 056005 (2019).
%
\bibitem{sarkar} A. Mukherjee, S. Ghosh, M. Mandal, P. Roy and S. Sarkar,
Phys. Rev. D {\bf 96}, 016024 (2017).
%
\bibitem{sarkar02} S. Ghosh, A. Mukherjee, M. Mandal, S. Sarkar and P. Roy,
Phys. Rev. D {\bf 96}, 116020 (2017).
%
\bibitem{zhang01} R. Zhang, W. Fu and Y. Liu, Eur. Phys. J. C {\bf 76} 307 (2016).
%
\bibitem{huan02} H. Liu, L. Yu and M. Huang, Phys. Rev. D {\bf 91}, 014017 (2015).
%
\bibitem{simonov01} M.~A.~Andreichikov and Y.~A.~Simonov,
Eur. Phys. J. C \textbf{78}, 902 (2018).
%
\bibitem{fraga01} G. Colucci, E .S. Fraga and A. Sedrakian, Phys. Lett. B {\bf 728}, 19 (2014)
%
\bibitem{aguirre} R. M. Aguirre
Phys. Rev. D {\bf 96}, 096013 (2017).
%
\bibitem{taya} H. Taya,
Phys. Rev. D {\bf 92}, 014038 (2015).
%
\bibitem{shinya} M. Kawaguchi and S. Matsuzaki,
Phys. Rev. D {\bf 93}, 125027(2016).
%
\bibitem{andersen01} J. O. Andersen, Phys. Rev. D {\bf 86}, 025020 (2012).
%
\bibitem{kojo} K. Hattori, T. Kojo and N. Su, Nucl. Phys. A {\bf 951}, 1 (2016).
%
\bibitem{simonov02} M. A. Andreichikov, B. O. Kerbikov, V. D. Orlovsky and Yu. A. Simonov, Phys. Rev. D {\bf 87}, 094029 (2013).
%
\bibitem{ghoshrho} S. Ghosh, A. Mukherjee, P. Roy, S. Sarkar, Phys. Rev. D {\bf 99}, 096004 (2019).
%
\bibitem{AMMmesons} N. Chaudhuri, S. Ghosh, S. Sarkar, P. Roy, Phys. Rev. D {\bf 99}, 116025 (2019).
%
\bibitem{TBspectralprop} S. Ghosh, A. Mukherjee, N. Chaudhuri, P. Roy and S. Sarkar,
Phys. Rev. D {\bf 101}, 056023 (2020).
%
\bibitem{ghoshEPJA} N. Chaudhuri, S. Ghosh, S. Sarkar, P. Roy, Eur. Phys. J. A {\bf 56}, 213 (2020).
%
\bibitem{bali02} G. S.  Bali, B. B. Brandt, G. Endr{\"o}di and B. Gl{\H o}ssle, Phys. Rev. D {\bf 97}, 034505 (2018). 
%
\bibitem{bali03} B.~B.~Brandt, G.~Bali, G.~Endrödi and B.~Glässle,
PoS LATTICE2015, 265 (2016).
%
\bibitem{luschev}  E. V. Luschevskaya, O.E. Solovjeva, O. E. Kochetkov and O. V. Teryaev, Nucl. Phys. B {\bf 898}, 627, (2015).
%
\bibitem{luschv02}  E. V. Luschevskaya, O. E. Kochetkov, O. V.Larina and O. V. Teryaev, Nucl. Phys. B{ \bf 884}, 1 (2014).
%
\bibitem{luschv03} E.V. Luschevskaya, O.E. Solovjeva and O.V. Teryaev, Phys. Lett. B {\bf 761}, 393 (2016).
%
\bibitem{hidaka} Y.~Hidaka and A.~Yamamoto, Phys.\ Rev.\ D {\bf 87}, 094502 (2013).
%
\bibitem{Ding2}H.~T.~Ding, S.~T.~Li, S.~Mukherjee, A.~Tomiya and X.~D.~Wang,
PoS LATTICE2019, 250 (2020).
%
\bibitem{Avila}
D.~Ávila and L.~Patiño,
Phys. Lett. B \textbf{795}, 689-693 (2019).
%
\bibitem{Avila:2020ved}
D.~Ávila and L.~Patiño,
J. High Energy Phys. \textbf{06}, 010 (2020)
%
\bibitem{dudal} N.~Callebaut, D.~Dudal and H.~Verschelde,
J.High Eenergy Phys. \textbf{03}, 033 (2013). 
%
\bibitem{dudal02} N.~Callebaut and D.~Dudal,
J. High Energy Phys. \textbf{01}, 055 (2014).
%
\bibitem{andrei02} M. A. Andreichikov, B. O. Kerbikov, V. D. Orlovsky and Yu. A. Simonov, Phys. Rev. D {\bf 89}, 074033 (2014).
%
\bibitem{he} B.-R. He, Phys. Rev. D {\bf 92}, 111503 (2015).
%
\bibitem{nucleon} A. Mukherjee, S. Ghosh, M. Mandal, S. Sarkar, P. Roy, Phys. Rev. D {\bf 98}, 056024 (2018).
%
\bibitem{barionslattice} G. Endr{\" o}di and G. Mark\'o, J. High Energy Phys. 08, 036 (2019). 

\bibitem{Ding}
H.-T. Ding, S.-T. Li, A. Tomiya, X.-D. Wang and Y. Zhang, arXiv:2008.00493 [hep-lat].

\bibitem{Das}
A. Das and N. Haque, Phys. Rev. D {\bf 101}, 074033  (2020).

\bibitem{Li}
J. Li, G. Cao and L. He, arXiv:2009.04697 [nucl-th].

\bibitem{pionmassmag}
A. Ayala, R. L. S. Farias, S. Hern\'andez Ortiz, L. A. Hern\'andez, D. Manreza Paret and R. Zamora, Phys. Rev. D {\bf 98}, 114008 (2018).

\bibitem{Scoccola}
M. Coppola, D. Gomez Dumm, S. Noguera and N. N. Scoccola, Phys. Rev. D {\bf 100}, 054014 (2019).

\end{thebibliography}
\end{document}